\documentclass[11pt,twocolumn,aps,prl,superscriptaddress,reprint,citeautoscript,floatfix]{revtex4-2}

\setcitestyle{super}
\usepackage[abs]{overpic}
\usepackage{graphicx,graphics}
\usepackage{dcolumn}
\usepackage{amsmath,amssymb,amsfonts}
\usepackage{gensymb}
\usepackage{latexsym,verbatim}
\usepackage{bm}
\usepackage{color}
\usepackage{ulem}
\usepackage{sidecap}
\usepackage[framemethod=default]{mdframed}
\usepackage[breaklinks=true,colorlinks=false,citecolor=blue,linkcolor=blue,urlcolor=blue]{hyperref}
\usepackage[makeroom]{cancel}
\usepackage{fancybox}
\usepackage{physics}
\usepackage{float}

\usepackage{gensymb}
\usepackage{wrapfig}

\usepackage[dvipsnames]{xcolor}
\definecolor{SIcolor}{RGB}{219, 48, 122}

\begin{document}
\author{Isabelle Y. Phinney}
\affiliation{Department of Chemistry and Chemical Biology, Harvard University, Cambridge, 02138, MA, USA.}
\author{Andrew Zimmerman}
\affiliation{Department of Physics, Harvard University, Cambridge, 02138, MA, USA.}
\author{Zeyu Hao}
\affiliation{Department of Physics, Harvard University, Cambridge, 02138, MA, USA.}
\author{Patrick J.~Ledwith}
\affiliation{Department of Physics, Harvard University, Cambridge, 02138, MA, USA.}
\author{Takashi Taniguchi}
\affiliation{Research Center for Materials Nanoarchitectonics, National Institute for Materials Science,  1-1 Namiki, Tsukuba 305-0044, Japan}
\author{Kenji Watanabe}
\affiliation{Research Center for Electronic and Optical Materials, National Institute for Materials Science, 1-1 Namiki, Tsukuba 305-0044, Japan}
\author{Ashvin Vishwanath}
\affiliation{Department of Physics, Harvard University, Cambridge, 02138, MA, USA.}
\author{Philip Kim}
\affiliation{Department of Physics, Harvard University, Cambridge, 02138, MA, USA.}

\title{Modulation of superconductivity across a Lifshitz transition in alternating-angle twisted quadrilayer graphene}

\begin{abstract}

We report electric field-controlled modulation of the Fermi surface topology and explore its effects on the superconducting state in alternating-angle twisted quadrilayer graphene (TQG). The unique combination of flat and dispersive bands in TQG allows us to simultaneously tune the band structure through carrier density, $n$, and displacement field, $D$.  From density-dependent Shubnikov-de Haas quantum oscillations and Hall measurements, we quantify the $D$-dependent bandwidth of the flat and dispersive bands and their hybridization. In the high $|D|$ regime, the increased bandwidth favors the single particle bands, which coincides exactly with the vanishing of the superconducting transition temperature $T_c$, showing that superconductivity in TQG is strongly bound to the symmetry-broken state. For a range of lower $|D|$ values, a Lifshitz transition occurs when the flat and dispersive band Fermi surfaces merge within the $\nu=+2$ symmetry-broken state. The superconducting state correspondingly shows an enhanced $T_c$, suggesting that the superconducting condensate is strongly dependent on the Fermi surface topology and density of states within this symmetry-broken state.

\end{abstract}
\maketitle


Since the discovery of superconductivity in twisted bilayer graphene (TBG), the field of twisted graphene, and, generally, moiré systems, has grown immensely\cite{cao2018sc, moirenatreview, andrei2020review, Balents2020review}. TBG offers gate- and twist angle-tunable narrow electronic bands, which easily lend themselves to the study of strongly correlated physics. Many interesting phases have been observed in these flat band systems, including superconductivity, correlated insulators, and ferromagnetism, all of which are likely tied to correlated electron interactions \cite{li2019, cao2018sc, cao2018correlated, yankowitz2019sc, sharpe2019, lu2019}. The superconducting phase has drawn particular interest because of the potential for an unconventional pairing mechanism. We seek to address how these broken symmetry states relate to the superconducting state, and how changes in Fermi surface topology and bandwidth affect the superconducting transition in this material. 

\begin{figure*}[htbp]
\centering
\begin{minipage}[t]{0.99 \textwidth}
\includegraphics[width=\textwidth]{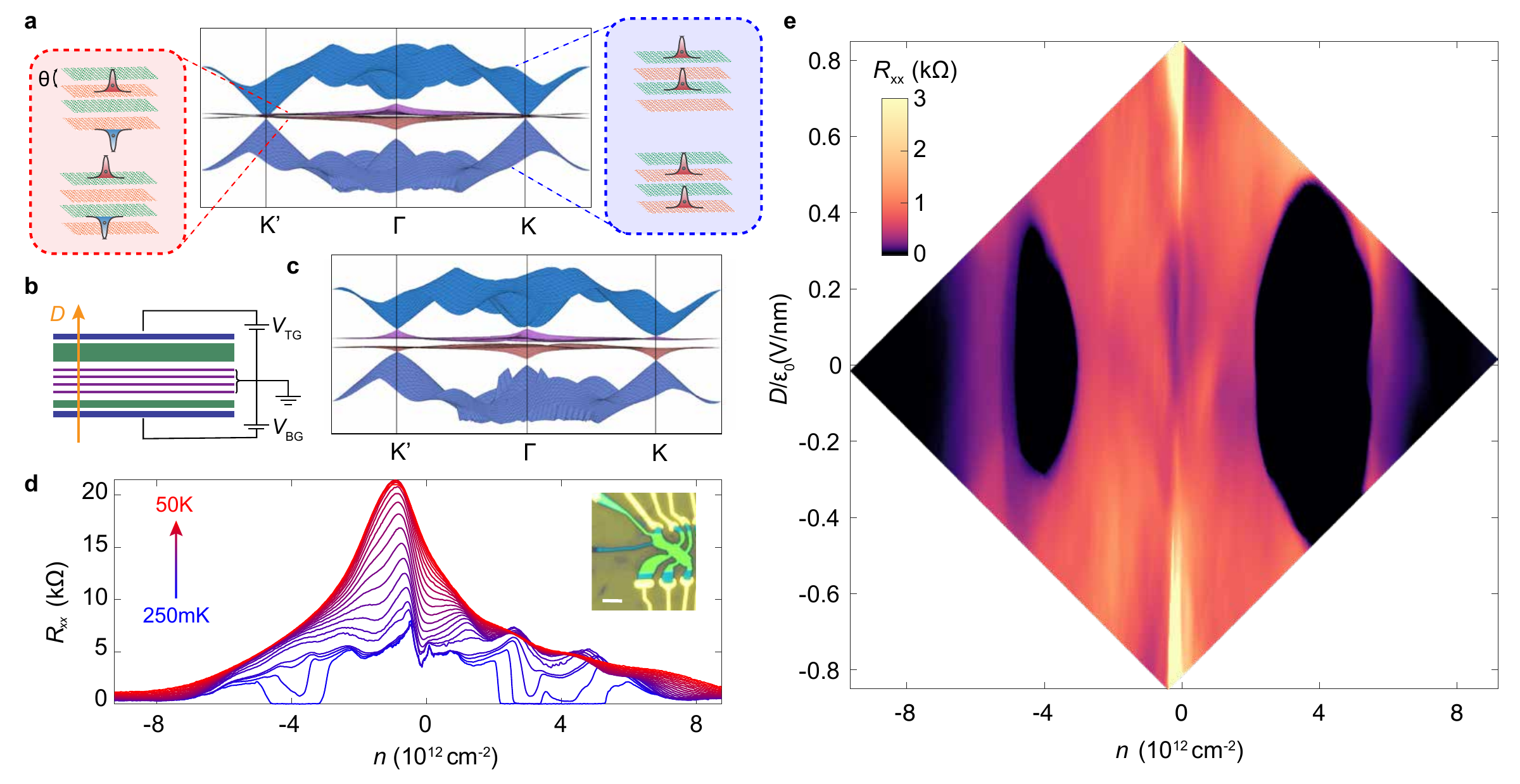}
	\caption{\textbf{a)} Single particle band structure of TQG at zero displacement field. Four layers of monolayer graphene are stacked with alternating twist angle (layers 1 and 3, green: $+\theta/2$, layers 2 and 4, orange: $-\theta/2$).  
    The $8\times 8$ Hamiltonian describing the tunneling between the four layers can be decomposed into two $4\times 4$ block-diagonal matrices, composed of the symmetric (antisymmetric) combinations of each layer's wavefunctions, as represented by the cartoon sketches. The tunneling in each block is scaled up (down) by a factor of $\phi$. Thus the two blocks directly describe two twisted bilayers with twist angle $\phi\theta$ and $\frac{1}{\phi}\theta$, i.e. a dispersive sector and a flat sector, respectively.     \textbf{b)} Schematic of dual graphite-gated TQG device. Graphite is shown in blue, graphene in purple, and insulating hexagonal boron nitride in green. The displacement field vector is defined as pointing from back gate to top gate. \textbf{c) } Single particle band structure at finite displacement field. \textbf{d) } Temperature dependence from 250~mK (blue) to 50~K (red) of TQG1, showing superconductivity emerging around 1.7~K near $\abs{n} = 4 \times 10^{12}$~cm${^{-2}}$ of the electron and hole sides. (Inset) Optical image of dual-encapsulated, dual-gated device. \textbf{e)} Four-terminal resistivity versus carrier density $n$ and displacement field $D$ at $T=250$\,mK. Two oblong pockets of superconductivity (black) emerge on both the electron and hole side, and are displacement field tunable.}
	\label{fig:Fig1}
 \end{minipage}	
\end{figure*}

Recently, the family of moiré graphene has been extended to include alternating angle twisted multilayer graphene (TMG), where electric field- and density-tunable superconductivity and correlated symmetry broken phases have been observed \cite{hao2021, park2021trilayer, park2022family, zhang2021ascendance, burg_emergence_2022}. 
The increased tunability of these transitions in TMG as compared to TBG is due to the presence of multiple bands near the Fermi level, which can be altered and hybridized by external electric fields. Near the magic angle, the strongly correlated behavior of TMG systems is characterized by flat bands (which we term the ``flat'' or ``magic sector'') that can be mapped to that of a twisted bilayer at the magic angle ($\approx 1.1\degree$)  \cite{khalaf2019}. 
Each additional pair of layers beyond the first two contributes a dispersive band that can be mapped directly to that of an off-magic-angle twisted bilayer (``dispersive sector''), while odd-layered TMG contains an additional monolayer-like Dirac cone (``dispersive cone''). This is illustrated in Figure \ref{fig:Fig1}a for four layers, with the magic sector in red and the dispersive sector in blue. Therefore, a systematic study of the electric field $D$ and density $n$ — both of which can be experimentally controlled via top and bottom electrostatic gates in the device (Figure \ref{fig:Fig1}b) — provides an opportunity to tune the single particle band structure and corresponding Fermi surface topology, enabling the exploration of their effect on superconductivity and the symmetry-broken correlated phases in TMG.

Despite their seemingly different low energy band structures, all TMG systems studied experimentally thus far exhibit very similar behavior, namely pockets of superconductivity near half filling of the conduction and valence band of the flat sector and symmetry breaking near quarter, half, and three-quarter fillings \cite{hao2021, park2021trilayer, park2022family, zhang_promotion_2022,burg_emergence_2022, kim2022}. Previous experimental works have intimated that the electron interaction-induced symmetry-broken state near half filling is necessary to the observed superconducting state, implying that the origin of superconductivity may be electron-electron mediated and that the pairing mechanism may require a specific isospin-polarized ground state \cite{cao2018sc, rozen2021, kerelsky2019, xie2019, jiang2019, choi2019stm, nuckolls2023}. However, a number of competing theories suggest alternative origins, including a more conventional electron-phonon mediated pairing \cite{polshyn2019, AllanPhononDriven, WuPhonondriven, Chen2024phonon, kwan2024kekulephonon}. 

In this work, we study alternating angle twisted quadrilayer graphene (TQG). Compared to twisted trilayer graphene (TTG), where a Dirac-like highly dispersive band is present, TQG offers less dispersive bands that can be readily tuned and hybridized with the flat band by a displacement field. We can broadly modulate the TQG band structure to explore the effects of bandwidth and Fermi surface size on symmetry-breaking and superconductivity, while using the dispersive band sector as a sensitive probe of the strongly interacting physics in the flat band sector.
Notably, away from charge neutrality, the Fermi surface of the magic sector is generally much larger than that of the dispersive sector.
When a displacement field, or out-of-plane electric field, $D$, is applied (Figure \ref{fig:Fig1}b), the flat and dispersive band sectors hybridize, resulting in single particle band structure changes as shown in Figure \ref{fig:Fig1}c as compared to figure \ref{fig:Fig1}a.


We fabricated three TQG devices, labeled TQG1, TQG2, and TQG3, all near $1.6^{\circ}$, which is slightly below the predicted magic angle of $1.77^{\circ}$\cite{khalaf2019, ledwith2021tb} (for detailed fabrication methods and device geometry see \hyperref[sec:SI]{Supplementary Information (S.I.)}). We focus primarily on TQG1 (device shown in Figure \ref{fig:Fig1}d inset) in the main text, with additional data presented in the \hyperref[sec:SI]{S.I.}. Figure \ref{fig:Fig1}d shows the measured longitudinal resistance $R_{xx}$ at  $D/\epsilon_0 = 0$~V/nm as a function of carrier density $n$, measured at different fixed temperatures. Superconductivity sets in around $1.7$~K near half-filling of the flat bands. 
Figure \ref{fig:Fig1}e shows the four-terminal resistance $R_{xx}$ versus carrier density $n$ and $D/\epsilon_0$. Superconductivity appears as black ovals on both the electron and hole side near half-filling of the flat band. 

The action of this out-of-plane electric field can be exemplified in the following way: in TBG at any angle, $D$-field will naively push charges to be more strongly localized on the top or bottom layer (depending on the sign of the charge), which in turn shifts the $k$-space Dirac-like cones at the mini-Brillouin zone corners $K_{\text{m}}$ and $K'_{\text{m}}$ in opposite directions. However, near the magic angle, the layers are strongly hybridized, so this layer polarization effect is negligible compared to the band hybridization, therefore displacement field has a weak effect. On the other hand, because TMG systems have one or more additional dispersive sectors on top of the flat band sector, displacement field can have a larger effect: it will be able to appreciably shift the weakly hybridized dispersive sector in opposite directions at $K_{\text{m}}$ and $K'_{\text{m}}$. Additionally, $D$ admits a tunneling term between the flat and dispersive sectors and thus hybridization between the two \cite{ledwith2021tb}. The latter can greatly change the overall band structure, for example, allowing the flat and dispersive sectors to hybridize near the $K_{\text{m}}$ and $K'_{\text{m}}$ points and open the transport gap experimentally visible at $n=0$ and $|D|\gtrsim 0.4$~V/nm in Figure \ref{fig:Fig1}e. This gap opening and hybridization can be seen in the band structure shown in Figure \ref{fig:Fig1}c. 

\begin{figure}[htbp]
\begin{minipage}[t]{0.49\textwidth}
	\centering
	\includegraphics[width=1\textwidth]{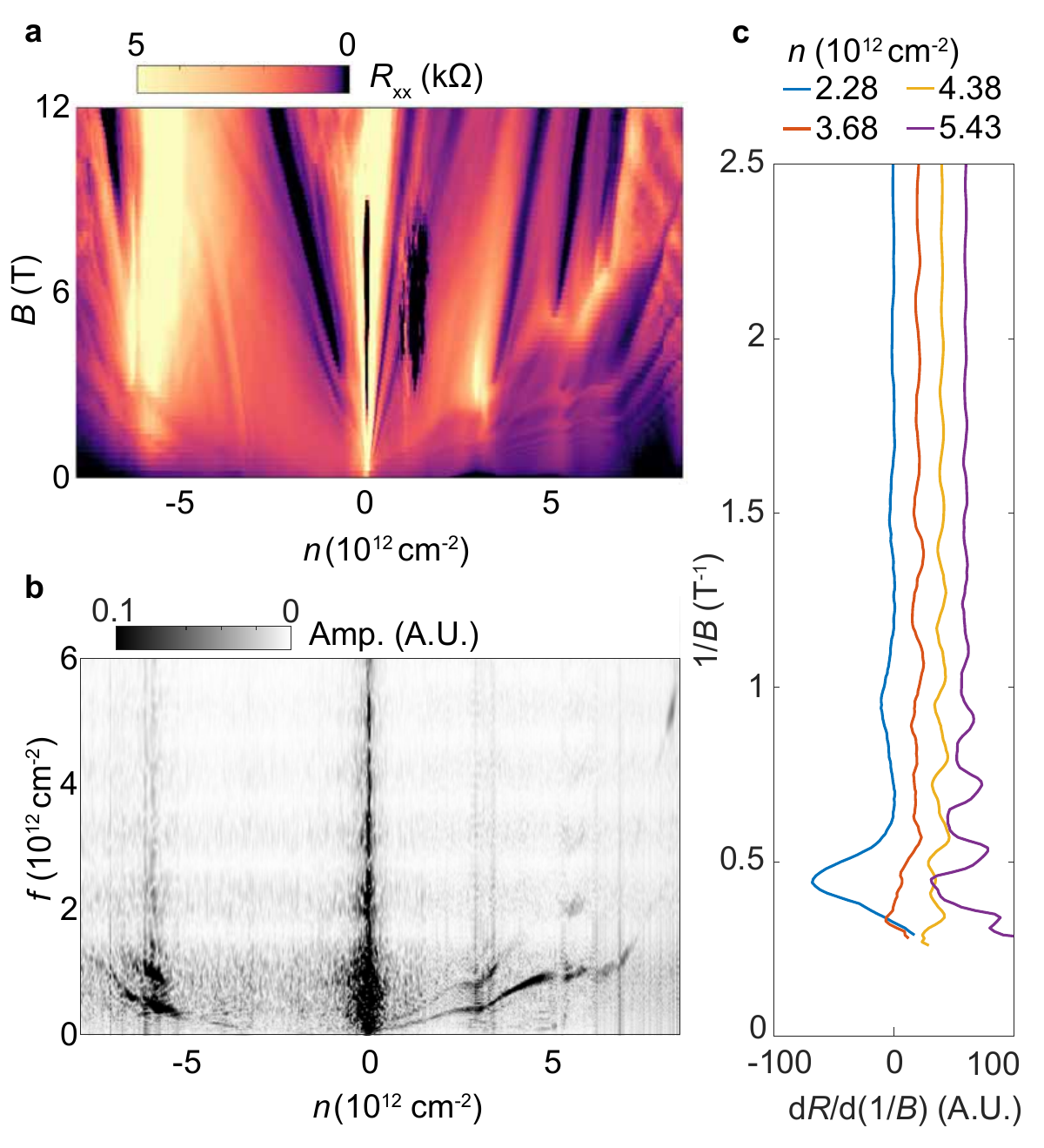}
	\caption{Both the flat and dispersive sectors appear in magnetotransport. \textbf{a)} $R_{xx}$ versus $B$ at $D/\epsilon_0 = 0$~V/nm. The flat band sector symmetry-breaks at integer fillings, as evidenced by zero resistance Chern insulator states. The dispersive sector's magnetooscillations manifest as a scalloped shape. \textbf{b)} Fast fourier transform (FFT) of (\textbf{a}) converted to units of $10^{12} cm^{-2}$ by Eq. \ref{eq:1}. The dispersive band appears as a dark line. \textbf{c) } Linecuts at various $n$ versus $1/B$, showing quantum oscillations due to the dispersive sector.}
	\label{fig:Fig2}
	\end{minipage}\hfill
\end{figure}

While the strongly-correlated flat sector physics dominates the transport, we can utilize the dispersive band as a sensitive probe to determine the carrier distribution between two bands. Magneto-oscillations provide a measure of the number and size of the corresponding Fermi surfaces active in each sector at any given filling, and are one of the primary probes we utilize to study the effect of hybridization and Fermi surface changes on symmetry-breaking and superconductivity in TQG. For example, from the $D/\epsilon_0=0$~V/nm Landau fan (LF) in Figure \ref{fig:Fig2}a, we observe Chern insulators at high field, resulting from the symmetry broken states of the flat bands (see \hyperref[sec:SI]{S.I.}). At the same time, oscillations at low field originate from the simultaneous filling of the dispersive sector, shown in Figure \ref{fig:Fig2}c. The periodicity of these Shubnikov de Haas oscillations (SdHO) gives the carrier density, $n_d$, in the dispersive subsystem:
\begin{equation}\label{eq:1}
    n_d = \frac{g e}{h}\frac{1}{\Delta (B^{-1})}
\end{equation}
where $\Delta (B^{-1})$ is the period of the SdHO in units of Tesla$^{-1}$, $e$ is the charge of an electron, $h$ is Planck's constant, and the degeneracy $g=8$ includes the layer degeneracy $K_\text{m}, K'_\text{m}$ \cite{zhang2019_lldegen}. The oscillations can be seen more clearly by taking a fast Fourier transform (FFT) of Figure \ref{fig:Fig2}a with respect to ${B^{-1}}$ and converting to the standard units of $n$, as shown in Figure \ref{fig:Fig2}b. Using Eq. \ref{eq:1} to calculate $n_d$ and the fact that the total carrier density will be the sum of the carriers in the flat and dispersive sectors, $n = n_f + n_d$, we find $\approx 15$\% of the carriers reside in the dispersive bands for $n > 0$ (electron). For $n < 0$ (hole), however, only $\approx 10$\% of the carriers reside in the dispersive bands. This electron-hole asymmetry must be accounted for in any theoretical description of the TQG band structure. 

It is convenient to use full-filling of the moiré unit cell as a metric, rather than the carrier density, so we will discuss our observations in terms of band filling $\nu = 4n/n_s$, where $n_0 = \sqrt{3} \theta^2/8a^2$ ($\theta$ the twist angle and $a$ the graphene lattice constant) is the total number of carriers required to fill one moiré unit cell. Because of the two or more co-existing sectors in twisted multilayer systems, it is also useful to define the filling of the flat sector alone:
\begin{equation}\label{eq:2}
    \nu_f =  4n_f /n_0 = 4(n-n_d)/n_0
\end{equation}
where $n_f \approx 0.85\, n$ for $n>0$ and $n_f \approx 0.9\, n$ for $n<0$. This is an important distinction to make as it is expected that the superconductivity primarily arises due to the flat band subsystem, rather than the dispersive subsystem. Indeed, many of the strongly correlated features in the data, including superconductivity, align with integer values of $\nu_f$ rather than $\nu$ as we will show below.

Since the two sectors are in equilibrium, the chemical potential of dispersive sector $\mu_d$, measured by the SdHO, can be used as a probe of the chemical potential $\mu_f$ of the flat band sector should be identical to equilibrium chemical potential $\mu$ and thereby extract the inverse compressibility $d\mu/dn$. We calculate $\mu$ using the linear dispersion of the low energy cones of the dispersive bands,
\begin{equation}\label{eq:3}
    \mu=\mu_d= v^*_F k_F = v^*_F\sqrt{\pi    n_d /2} 
\end{equation}
assuming a reduced Fermi velocity $v^*_F = 0.5 \times 10^6$~m/s.

Drawing a direct comparison to Refs. \cite{zondiner2020_cascade,park2020hund,saito2021isospin,rozen2021}, we can interpret the vanishing or even negative slope of $d\mu/dn$ near integer fillings as symmetry-breaking phase transitions. A more detailed discussion is included in the \hyperref[sec:SI]{S.I.}.

\begin{figure}[htbp]
\begin{minipage}[t]{0.49\textwidth}
	\centering
	\includegraphics[width=1\textwidth]{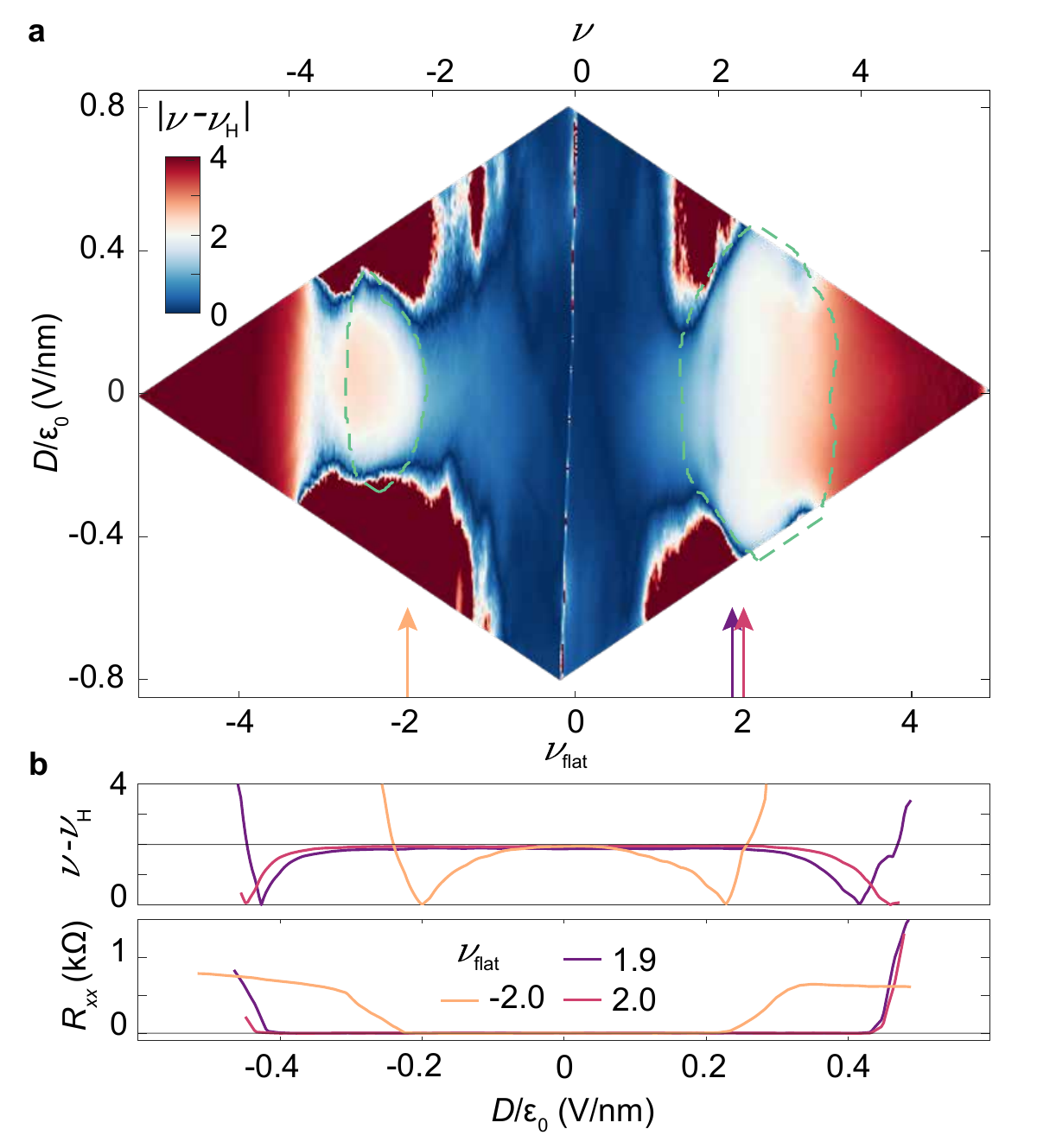}
	\caption{\textbf{a)} Subtracted Hall density, $\nu-\nu_H$, anti-symmetrized at $\pm 0.5$~T. The large white plateaus indicate the $\nu=2$ symmetry broken regime where superconductivity appears at zero field. \textbf{b)} Line cuts at various fillings near $\nu_{\text{f}} = 2$ versus $D$ comparing subtracted Hall density (upper) to $R_{xx}$ (lower), showing that superconductivity aligns well with the symmetry-broken state at $\nu=2$.}
	\label{fig:Fig3}
\end{minipage} \hfill
\end{figure}

Symmetry-breaking features are also easily visualized by considering the subtracted Hall density, $\nu-\nu_H$.\cite{saito2021isospin} Figure \ref{fig:Fig3}a plots this quantity over the maximum range of $D$ and $n$ accessible in TQG1. The symmetry broken state at half filling, for example, appears as a broad white plateau of value $|\nu-\nu_H| = 2$ on both the electron and hole sides. Even more telling is how closely the region of superconductivity matches these boundaries (dashed line fitted from Figure \ref{fig:Fig1}e), showing that superconductivity in TQG is strongly tied to the symmetry-broken state at $\nu_f=2$. Figure \ref{fig:Fig3}b and c show corresponding line cuts of $\nu-\nu_H$ and $R_{xx}$ to further emphasize this point. Notably, the system appears to favor the flavor-polarized ground state of $\nu_f=2$ and symmetry-breaks to this state even before reaching $\nu_f=2$, presumably meaning the carriers in the flat bands will be hole-like in this small region. The superconductivity correspondingly extends over this entire region.

From the subtracted Hall density, we can see there are regions of clear symmetry-breaking but also regions that follow a simpler single particle picture. For example, starting above $|D| \approx 0.4$~V/nm, moving from charge neutrality outwards, we see the system does not symmetry break but rather maintains the single particle band structure, switching directly from $\nu-\nu_H = 0$ to $\nu-\nu_H = 4$ across the single particle van Hove singularity (vHS). In contrast, for $|D| \lesssim 0.4$~V/nm, the system symmetry-breaks at each integer filling $\nu_f$ between 1 and 4, as indicated by the four different colors for each flavor in Figure \ref{fig:Fig3}a. We illustrate these boundaries between single-particle and flavor polarized regimes in with a cartoon in Supplementary Figure \ref{fig:Fig5SI}. The superconductivity, indicated by the dashed green line in Figure \ref{fig:Fig3}a, is clearly bounded above and below by the single-particle regime, further supporting the claim that the physics of the symmetry-broken $\nu_f=2$ state is vitally tied to the origin of the superconducting state. Small overlaps of the superconducting regime and the single particle regime are due to magnetic field band reconstruction at $0.5$~T.

\begin{figure}[htbp]
\begin{minipage}[t]{0.49\textwidth}
	\centering
	\includegraphics[width=1\textwidth]{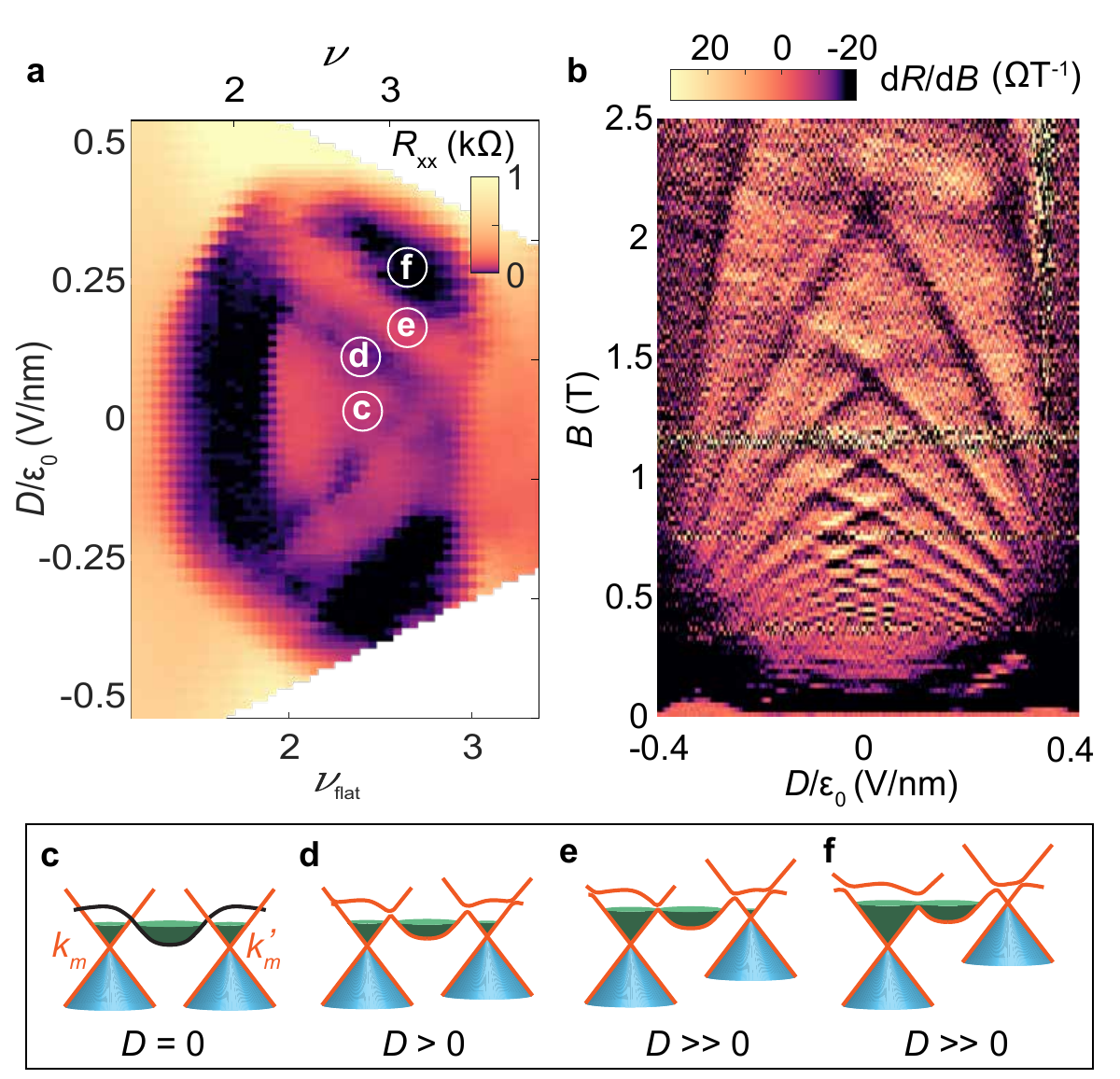}
	\caption{Magnetotransport showing a Lifshitz transition occurring at high displacement field. \textbf{a)} Electron-side superconducting dome at $B_{\perp}=50$~mT. Two regions of stronger superconductivity symmetric in displacement field are visible as black, zero-resistance states, where the superconductivity is not yet suppressed by $B$. \textbf{b)} $dR/dB$ showing quantum oscillations in the dispersive band versus displacement field at $\nu_{\text{f}} = 2.4$. The derivative simply removes a background offset. \textbf{c-f)} Cartoon showing the flat band and dispersive bands as a function of displacement field and filling.} 
	\label{fig:Fig4}
\end{minipage}
\end{figure}

One unusual feature of the superconducting dome observed in all three devices is the presence of distinct regions of stronger and weaker superconductivity. 
Figure \ref{fig:Fig4}a shows $R_{xx}$ measured near the electron side superconducting regime at $\nu_f=2$ as a function of $\nu_f$ and $D$. The measurement was performed at the base temperature $T=250$~mK but with a small constant magnetic field $B_{\perp} = 50$~mT to weaken the superconductivity. 
Within the superconducting oval region in the $\nu_f$ and $D$ plane, the stronger (weaker) superconducting region appears as lower (higher) resistance state with darker (lighter) color coding. Notably, we observe three stronger superconducting regions: one region at $\nu_f<2$ and two regions at $\nu_f>2$ at high $D$.  
There are also two characteristic triangular regions of weak superconductivity separated slightly stronger superconductivity regime between them. These weaker superconducting regime consistently appeared in all three devices and can be easily suppressed by increasing $T$, applying $B_\perp$ or applying a bias current $I$ (see the additional characterization of the superconducting dome in $n$ and $D$ in the \hyperref[sec:SI]{S.I.} versus $T$, $B_{\perp}$, and $B_{\parallel}$). 

In order to have a better understanding of the band structure in this regime and how it might relate to the varying strength of superconductivity, we use the same SdHO magnetotransport technique of the dispersive bands described earlier to measure the Fermi surface sizes. We attempt to correlate features observed in the superconductivity with the Fermi surface and band structure.
Figure \ref{fig:Fig4}b shows $dR_{xx}/dB$ as a function of $D$ and $B$. Near $D\approx 0$~V/nm, we find two sets of SdH oscillations intersect, suggesting the presence of two same sized Fermi surfaces at $D=0$~V/nm. As $|D|$ increases from the zero value, the degeneracy of SdH oscillations is lifted, and one SdH oscillation becomes faster and the other becomes slower, suggesting two Fermi surface resiponsible for SdH now have different sizes.

In order for more quantitative analysis, we take an FFT of the SdHO to obtain $n_d$ (S.I. Figure \ref{fig:Fig6SI}). Assuming a linear dispersion, we calculate the distance between the Fermi level and the Dirac point of the two cones in meV, plotted in Supplementary Figure \ref{fig:Fig6SI} versus $D$ at two different carrier densities $\nu_f$. At $D/\epsilon_0=0$~V/nm, the two cones are equally spaced in energy and their Fermi surfaces are the same size (Figure \ref{fig:Fig4}c). However, as we apply $D$, two sets of Fermi surfaces appear, one of which moves down with $D>0$ and the other moves in the opposite direction (Figure \ref{fig:Fig4}d-f). In the FFT (Supplementary Figure \ref{fig:Fig6SI}), the two Fermi surfaces appear to ``cross'' at $D/\epsilon_0=0$~V/nm. To first order, we know $D$ couples linearly to the dispersive subsector, acting with opposite sign at $K_m$ and $K'_m$, the mini-Brillouin zone valleys. This will push the cone at $K_m$ up while pushing the cone at $K'_m$ down in energy, and vice versa for opposite $D$. The observed difference in energy ($\approx 5$~meV at $D/\epsilon_0=0.3$\,V/nm) matches that predicted by theory \cite{carr2020, ledwith2021tb}.

We also note the cone that is pushed down in energy (therefore appearing as larger Fermi surface or higher energy branch in Supplementary Figure \ref{fig:Fig6SI}) seems to disappear at smaller $D$ as $\nu_f$ increases. With finite displacement field applied, we expect the flat bands and dispersive bands to hybridize near where they cross in energy, as shown in the cartoon in Figure \ref{fig:Fig4}f. When the lower cone's Fermi surface ``vanishes'', it is simply hybridizing with the flat band (Figure \ref{fig:Fig4}f), and the system undergoes a Lifshitz transition as the two Fermi surfaces merge into a single larger one. The second strongest region of superconductivity seems to occur exactly after this point: when one of the cones hybridizes with the flat band and the Fermi level is high enough to encircle this new larger Fermi surface (Figure. \ref{fig:Fig4}f). 
The close connection between SdHO and the strength of superconductivity in TQG indicates that the graphene moiré superconductivity can be very sensitive to single particle bands, in addition to the commonly seen connection with symmetry breaking phase.

In conclusion, we demonstrate how the unique combination of flat and dispersive bands in TQG simultaneously tune the band structure through carrier density and displacement field. We find the emergence of the superconductivity is strongly tied to the $\nu=\pm2$ symmetry-broken state and the hybridization between the dispersive and flat bands results in a Lifshitz transition. The superconductivity is found to be enhanced across this transition in the high density of states regime where the flat and dispersive band Fermi surface have merged. Overall, we show the high degree of tunability in TQG paves a route for understanding the relationship between symmetry-breaking, band structure, and superconductivity in multilayer moiré graphene systems.

We acknowledge helpful discussions with Maine Christos and Eslam Khalaf. The major experimental work is supported by DOE (DE- SC0012260). P.K. acknowledges support from ARO MUIR (W911NF-21-2-0147). A.V. acknowledges funding support from NSF-DMR 2220703. I.Y.P acknowledges the support of the Herchel-Smith fellowship, the NSF Graduate Research Fellowships Program, and the Hertz Foundation. K.W. and T.T. acknowledge support from the JSPS KAKENHI (Grant Numbers 21H05233 and 23H02052) , the CREST (JPMJCR24A5), JST and World Premier International Research Center Initiative (WPI), MEXT, Japan. Nanofabrication was performed at the Center for Nanoscale Systems at Harvard, supported in part by an NSF NNIN award ECS-00335765.




\bibliography{MAINNotes.bib}

\newcommand{\noopsort}[1]{} \newcommand{\printfirst}[2]{#1}
  \newcommand{\singleletter}[1]{#1} \newcommand{\switchargs}[2]{#2#1}
\begin{thebibliography}{40}%
\makeatletter
\providecommand \@ifxundefined [1]{%
 \@ifx{#1\undefined}
}%
\providecommand \@ifnum [1]{%
 \ifnum #1\expandafter \@firstoftwo
 \else \expandafter \@secondoftwo
 \fi
}%
\providecommand \@ifx [1]{%
 \ifx #1\expandafter \@firstoftwo
 \else \expandafter \@secondoftwo
 \fi
}%
\providecommand \natexlab [1]{#1}%
\providecommand \enquote  [1]{``#1''}%
\providecommand \bibnamefont  [1]{#1}%
\providecommand \bibfnamefont [1]{#1}%
\providecommand \citenamefont [1]{#1}%
\providecommand \href@noop [0]{\@secondoftwo}%
\providecommand \href [0]{\begingroup \@sanitize@url \@href}%
\providecommand \@href[1]{\@@startlink{#1}\@@href}%
\providecommand \@@href[1]{\endgroup#1\@@endlink}%
\providecommand \@sanitize@url [0]{\catcode `\\12\catcode `\$12\catcode
  `\&12\catcode `\#12\catcode `\^12\catcode `\_12\catcode `\%12\relax}%
\providecommand \@@startlink[1]{}%
\providecommand \@@endlink[0]{}%
\providecommand \url  [0]{\begingroup\@sanitize@url \@url }%
\providecommand \@url [1]{\endgroup\@href {#1}{\urlprefix }}%
\providecommand \urlprefix  [0]{URL }%
\providecommand \Eprint [0]{\href }%
\providecommand \doibase [0]{https://doi.org/}%
\providecommand \selectlanguage [0]{\@gobble}%
\providecommand \bibinfo  [0]{\@secondoftwo}%
\providecommand \bibfield  [0]{\@secondoftwo}%
\providecommand \translation [1]{[#1]}%
\providecommand \BibitemOpen [0]{}%
\providecommand \bibitemStop [0]{}%
\providecommand \bibitemNoStop [0]{.\EOS\space}%
\providecommand \EOS [0]{\spacefactor3000\relax}%
\providecommand \BibitemShut  [1]{\csname bibitem#1\endcsname}%
\let\auto@bib@innerbib\@empty
\bibitem [{\citenamefont {{Cao}}\ \emph
  {et~al.}(2018{\natexlab{a}})\citenamefont {{Cao}}, \citenamefont {{Fatemi}},
  \citenamefont {{Fang}}, \citenamefont {{Watanabe}}, \citenamefont
  {{Taniguchi}}, \citenamefont {{Kaxiras}},\ and\ \citenamefont
  {{Jarillo-Herrero}}}]{cao2018sc}%
  \BibitemOpen
  \bibfield  {author} {\bibinfo {author} {\bibfnamefont {Y.}~\bibnamefont
  {{Cao}}}, \bibinfo {author} {\bibfnamefont {V.}~\bibnamefont {{Fatemi}}},
  \bibinfo {author} {\bibfnamefont {S.}~\bibnamefont {{Fang}}}, \bibinfo
  {author} {\bibfnamefont {K.}~\bibnamefont {{Watanabe}}}, \bibinfo {author}
  {\bibfnamefont {T.}~\bibnamefont {{Taniguchi}}}, \bibinfo {author}
  {\bibfnamefont {E.}~\bibnamefont {{Kaxiras}}},\ and\ \bibinfo {author}
  {\bibfnamefont {P.}~\bibnamefont {{Jarillo-Herrero}}},\ }\bibfield  {title}
  {\bibinfo {title} {{Unconventional superconductivity in magic-angle graphene
  superlattices}},\ }\href {https://doi.org/10.1038/nature26160} {\bibfield
  {journal} {\bibinfo  {journal} {\nat}\ }\textbf {\bibinfo {volume} {556}},\
  \bibinfo {pages} {43} (\bibinfo {year} {2018}{\natexlab{a}})},\ \Eprint
  {https://arxiv.org/abs/1803.02342} {arXiv:1803.02342 [cond-mat.mes-hall]}
  \BibitemShut {NoStop}%
\bibitem [{moi(2021)}]{moirenatreview}%
  \BibitemOpen
  \bibfield  {title} {\bibinfo {title} {{Moir{\'e} magic three years on}},\
  }\href {https://doi.org/10.1038/s41578-021-00298-9} {\bibfield  {journal}
  {\bibinfo  {journal} {Nature Reviews Materials}\ }\textbf {\bibinfo {volume}
  {6}},\ \bibinfo {pages} {191} (\bibinfo {year} {2021})}\BibitemShut {NoStop}%
\bibitem [{\citenamefont {{Andrei}}\ and\ \citenamefont
  {{MacDonald}}(2020)}]{andrei2020review}%
  \BibitemOpen
  \bibfield  {author} {\bibinfo {author} {\bibfnamefont {E.~Y.}\ \bibnamefont
  {{Andrei}}}\ and\ \bibinfo {author} {\bibfnamefont {A.~H.}\ \bibnamefont
  {{MacDonald}}},\ }\bibfield  {title} {\bibinfo {title} {{Graphene bilayers
  with a twist}},\ }\href {https://doi.org/10.1038/s41563-020-00840-0}
  {\bibfield  {journal} {\bibinfo  {journal} {Nature Materials}\ }\textbf
  {\bibinfo {volume} {19}},\ \bibinfo {pages} {1265} (\bibinfo {year}
  {2020})},\ \Eprint {https://arxiv.org/abs/2008.08129} {arXiv:2008.08129
  [cond-mat.mes-hall]} \BibitemShut {NoStop}%
\bibitem [{\citenamefont {Balents}\ \emph {et~al.}(2020)\citenamefont
  {Balents}, \citenamefont {Dean}, \citenamefont {Efetov},\ and\ \citenamefont
  {Young}}]{Balents2020review}%
  \BibitemOpen
  \bibfield  {author} {\bibinfo {author} {\bibfnamefont {L.}~\bibnamefont
  {Balents}}, \bibinfo {author} {\bibfnamefont {C.~R.}\ \bibnamefont {Dean}},
  \bibinfo {author} {\bibfnamefont {D.~K.}\ \bibnamefont {Efetov}},\ and\
  \bibinfo {author} {\bibfnamefont {A.~F.}\ \bibnamefont {Young}},\ }\bibfield
  {title} {\bibinfo {title} {Superconductivity and strong correlations in
  moir{\'e} flat bands},\ }\href {https://doi.org/10.1038/s41567-020-0906-9}
  {\bibfield  {journal} {\bibinfo  {journal} {Nat. Phys.}\ }\textbf {\bibinfo
  {volume} {16}},\ \bibinfo {pages} {725} (\bibinfo {year} {2020})}\BibitemShut
  {NoStop}%
\bibitem [{\citenamefont {{Li}}\ \emph {et~al.}(2019)\citenamefont {{Li}},
  \citenamefont {{Wu}},\ and\ \citenamefont {{Das Sarma}}}]{li2019}%
  \BibitemOpen
  \bibfield  {author} {\bibinfo {author} {\bibfnamefont {X.}~\bibnamefont
  {{Li}}}, \bibinfo {author} {\bibfnamefont {F.}~\bibnamefont {{Wu}}},\ and\
  \bibinfo {author} {\bibfnamefont {S.}~\bibnamefont {{Das Sarma}}},\
  }\bibfield  {title} {\bibinfo {title} {{Phonon scattering induced carrier
  resistivity in twisted double bilayer graphene}},\ }\href@noop {} {\bibfield
  {journal} {\bibinfo  {journal} {arXiv e-prints}\ ,\ \bibinfo {eid}
  {arXiv:1906.08224}} (\bibinfo {year} {2019})},\ \Eprint
  {https://arxiv.org/abs/1906.08224} {arXiv:1906.08224 [cond-mat.str-el]}
  \BibitemShut {NoStop}%
\bibitem [{\citenamefont {{Cao}}\ \emph
  {et~al.}(2018{\natexlab{b}})\citenamefont {{Cao}}, \citenamefont {{Fatemi}},
  \citenamefont {{Demir}}, \citenamefont {{Fang}}, \citenamefont {{Tomarken}},
  \citenamefont {{Luo}}, \citenamefont {{Sanchez-Yamagishi}}, \citenamefont
  {{Watanabe}}, \citenamefont {{Taniguchi}}, \citenamefont {{Kaxiras}},
  \citenamefont {{Ashoori}},\ and\ \citenamefont
  {{Jarillo-Herrero}}}]{cao2018correlated}%
  \BibitemOpen
  \bibfield  {author} {\bibinfo {author} {\bibfnamefont {Y.}~\bibnamefont
  {{Cao}}}, \bibinfo {author} {\bibfnamefont {V.}~\bibnamefont {{Fatemi}}},
  \bibinfo {author} {\bibfnamefont {A.}~\bibnamefont {{Demir}}}, \bibinfo
  {author} {\bibfnamefont {S.}~\bibnamefont {{Fang}}}, \bibinfo {author}
  {\bibfnamefont {S.~L.}\ \bibnamefont {{Tomarken}}}, \bibinfo {author}
  {\bibfnamefont {J.~Y.}\ \bibnamefont {{Luo}}}, \bibinfo {author}
  {\bibfnamefont {J.~D.}\ \bibnamefont {{Sanchez-Yamagishi}}}, \bibinfo
  {author} {\bibfnamefont {K.}~\bibnamefont {{Watanabe}}}, \bibinfo {author}
  {\bibfnamefont {T.}~\bibnamefont {{Taniguchi}}}, \bibinfo {author}
  {\bibfnamefont {E.}~\bibnamefont {{Kaxiras}}}, \bibinfo {author}
  {\bibfnamefont {R.~C.}\ \bibnamefont {{Ashoori}}},\ and\ \bibinfo {author}
  {\bibfnamefont {P.}~\bibnamefont {{Jarillo-Herrero}}},\ }\bibfield  {title}
  {\bibinfo {title} {{Correlated insulator behaviour at half-filling in
  magic-angle graphene superlattices}},\ }\href
  {https://doi.org/10.1038/nature26154} {\bibfield  {journal} {\bibinfo
  {journal} {Nature}\ }\textbf {\bibinfo {volume} {556}},\ \bibinfo {pages}
  {80} (\bibinfo {year} {2018}{\natexlab{b}})},\ \Eprint
  {https://arxiv.org/abs/1802.00553} {arXiv:1802.00553 [cond-mat.mes-hall]}
  \BibitemShut {NoStop}%
\bibitem [{\citenamefont {{Yankowitz}}\ \emph {et~al.}(2019)\citenamefont
  {{Yankowitz}}, \citenamefont {{Chen}}, \citenamefont {{Polshyn}},
  \citenamefont {{Zhang}}, \citenamefont {{Watanabe}}, \citenamefont
  {{Taniguchi}}, \citenamefont {{Graf}}, \citenamefont {{Young}},\ and\
  \citenamefont {{Dean}}}]{yankowitz2019sc}%
  \BibitemOpen
  \bibfield  {author} {\bibinfo {author} {\bibfnamefont {M.}~\bibnamefont
  {{Yankowitz}}}, \bibinfo {author} {\bibfnamefont {S.}~\bibnamefont {{Chen}}},
  \bibinfo {author} {\bibfnamefont {H.}~\bibnamefont {{Polshyn}}}, \bibinfo
  {author} {\bibfnamefont {Y.}~\bibnamefont {{Zhang}}}, \bibinfo {author}
  {\bibfnamefont {K.}~\bibnamefont {{Watanabe}}}, \bibinfo {author}
  {\bibfnamefont {T.}~\bibnamefont {{Taniguchi}}}, \bibinfo {author}
  {\bibfnamefont {D.}~\bibnamefont {{Graf}}}, \bibinfo {author} {\bibfnamefont
  {A.~F.}\ \bibnamefont {{Young}}},\ and\ \bibinfo {author} {\bibfnamefont
  {C.~R.}\ \bibnamefont {{Dean}}},\ }\bibfield  {title} {\bibinfo {title}
  {{Tuning superconductivity in twisted bilayer graphene}},\ }\href
  {https://doi.org/10.1126/science.aav1910} {\bibfield  {journal} {\bibinfo
  {journal} {Science}\ }\textbf {\bibinfo {volume} {363}},\ \bibinfo {pages}
  {1059} (\bibinfo {year} {2019})},\ \Eprint {https://arxiv.org/abs/1808.07865}
  {arXiv:1808.07865 [cond-mat.mes-hall]} \BibitemShut {NoStop}%
\bibitem [{\citenamefont {{Sharpe}}\ \emph {et~al.}(2019)\citenamefont
  {{Sharpe}}, \citenamefont {{Fox}}, \citenamefont {{Barnard}}, \citenamefont
  {{Finney}}, \citenamefont {{Watanabe}}, \citenamefont {{Taniguchi}},
  \citenamefont {{Kastner}},\ and\ \citenamefont
  {{Goldhaber-Gordon}}}]{sharpe2019}%
  \BibitemOpen
  \bibfield  {author} {\bibinfo {author} {\bibfnamefont {A.~L.}\ \bibnamefont
  {{Sharpe}}}, \bibinfo {author} {\bibfnamefont {E.~J.}\ \bibnamefont {{Fox}}},
  \bibinfo {author} {\bibfnamefont {A.~W.}\ \bibnamefont {{Barnard}}}, \bibinfo
  {author} {\bibfnamefont {J.}~\bibnamefont {{Finney}}}, \bibinfo {author}
  {\bibfnamefont {K.}~\bibnamefont {{Watanabe}}}, \bibinfo {author}
  {\bibfnamefont {T.}~\bibnamefont {{Taniguchi}}}, \bibinfo {author}
  {\bibfnamefont {M.~A.}\ \bibnamefont {{Kastner}}},\ and\ \bibinfo {author}
  {\bibfnamefont {D.}~\bibnamefont {{Goldhaber-Gordon}}},\ }\bibfield  {title}
  {\bibinfo {title} {{Emergent ferromagnetism near three-quarters filling in
  twisted bilayer graphene}},\ }\href {https://doi.org/10.1126/science.aaw3780}
  {\bibfield  {journal} {\bibinfo  {journal} {Science}\ }\textbf {\bibinfo
  {volume} {365}},\ \bibinfo {pages} {605} (\bibinfo {year} {2019})},\ \Eprint
  {https://arxiv.org/abs/1901.03520} {arXiv:1901.03520 [cond-mat.mes-hall]}
  \BibitemShut {NoStop}%
\bibitem [{\citenamefont {{Lu}}\ \emph {et~al.}(2019)\citenamefont {{Lu}},
  \citenamefont {{Stepanov}}, \citenamefont {{Yang}}, \citenamefont {{Xie}},
  \citenamefont {{Aamir}}, \citenamefont {{Das}}, \citenamefont {{Urgell}},
  \citenamefont {{Watanabe}}, \citenamefont {{Taniguchi}}, \citenamefont
  {{Zhang}}, \citenamefont {{Bachtold}}, \citenamefont {{MacDonald}},\ and\
  \citenamefont {{Efetov}}}]{lu2019}%
  \BibitemOpen
  \bibfield  {author} {\bibinfo {author} {\bibfnamefont {X.}~\bibnamefont
  {{Lu}}}, \bibinfo {author} {\bibfnamefont {P.}~\bibnamefont {{Stepanov}}},
  \bibinfo {author} {\bibfnamefont {W.}~\bibnamefont {{Yang}}}, \bibinfo
  {author} {\bibfnamefont {M.}~\bibnamefont {{Xie}}}, \bibinfo {author}
  {\bibfnamefont {M.~A.}\ \bibnamefont {{Aamir}}}, \bibinfo {author}
  {\bibfnamefont {I.}~\bibnamefont {{Das}}}, \bibinfo {author} {\bibfnamefont
  {C.}~\bibnamefont {{Urgell}}}, \bibinfo {author} {\bibfnamefont
  {K.}~\bibnamefont {{Watanabe}}}, \bibinfo {author} {\bibfnamefont
  {T.}~\bibnamefont {{Taniguchi}}}, \bibinfo {author} {\bibfnamefont
  {G.}~\bibnamefont {{Zhang}}}, \bibinfo {author} {\bibfnamefont
  {A.}~\bibnamefont {{Bachtold}}}, \bibinfo {author} {\bibfnamefont {A.~H.}\
  \bibnamefont {{MacDonald}}},\ and\ \bibinfo {author} {\bibfnamefont {D.~K.}\
  \bibnamefont {{Efetov}}},\ }\bibfield  {title} {\bibinfo {title}
  {{Superconductors, orbital magnets and correlated states in magic-angle
  bilayer graphene}},\ }\href {https://doi.org/10.1038/s41586-019-1695-0}
  {\bibfield  {journal} {\bibinfo  {journal} {\nat}\ }\textbf {\bibinfo
  {volume} {574}},\ \bibinfo {pages} {653} (\bibinfo {year} {2019})},\ \Eprint
  {https://arxiv.org/abs/1903.06513} {arXiv:1903.06513 [cond-mat.str-el]}
  \BibitemShut {NoStop}%
\bibitem [{\citenamefont {{Hao}}\ \emph {et~al.}(2021)\citenamefont {{Hao}},
  \citenamefont {{Zimmerman}}, \citenamefont {{Ledwith}}, \citenamefont
  {{Khalaf}}, \citenamefont {{Najafabadi}}, \citenamefont {{Watanabe}},
  \citenamefont {{Taniguchi}}, \citenamefont {{Vishwanath}},\ and\
  \citenamefont {{Kim}}}]{hao2021}%
  \BibitemOpen
  \bibfield  {author} {\bibinfo {author} {\bibfnamefont {Z.}~\bibnamefont
  {{Hao}}}, \bibinfo {author} {\bibfnamefont {A.~M.}\ \bibnamefont
  {{Zimmerman}}}, \bibinfo {author} {\bibfnamefont {P.}~\bibnamefont
  {{Ledwith}}}, \bibinfo {author} {\bibfnamefont {E.}~\bibnamefont {{Khalaf}}},
  \bibinfo {author} {\bibfnamefont {D.~H.}\ \bibnamefont {{Najafabadi}}},
  \bibinfo {author} {\bibfnamefont {K.}~\bibnamefont {{Watanabe}}}, \bibinfo
  {author} {\bibfnamefont {T.}~\bibnamefont {{Taniguchi}}}, \bibinfo {author}
  {\bibfnamefont {A.}~\bibnamefont {{Vishwanath}}},\ and\ \bibinfo {author}
  {\bibfnamefont {P.}~\bibnamefont {{Kim}}},\ }\bibfield  {title} {\bibinfo
  {title} {{Electric field{\textendash}tunable superconductivity in
  alternating-twist magic-angle trilayer graphene}},\ }\href
  {https://doi.org/10.1126/science.abg0399} {\bibfield  {journal} {\bibinfo
  {journal} {Science}\ }\textbf {\bibinfo {volume} {371}},\ \bibinfo {pages}
  {1133} (\bibinfo {year} {2021})},\ \Eprint {https://arxiv.org/abs/2012.02773}
  {arXiv:2012.02773 [cond-mat.supr-con]} \BibitemShut {NoStop}%
\bibitem [{\citenamefont {{Park}}\ \emph {et~al.}(2021)\citenamefont {{Park}},
  \citenamefont {{Cao}}, \citenamefont {{Watanabe}}, \citenamefont
  {{Taniguchi}},\ and\ \citenamefont {{Jarillo-Herrero}}}]{park2021trilayer}%
  \BibitemOpen
  \bibfield  {author} {\bibinfo {author} {\bibfnamefont {J.~M.}\ \bibnamefont
  {{Park}}}, \bibinfo {author} {\bibfnamefont {Y.}~\bibnamefont {{Cao}}},
  \bibinfo {author} {\bibfnamefont {K.}~\bibnamefont {{Watanabe}}}, \bibinfo
  {author} {\bibfnamefont {T.}~\bibnamefont {{Taniguchi}}},\ and\ \bibinfo
  {author} {\bibfnamefont {P.}~\bibnamefont {{Jarillo-Herrero}}},\ }\bibfield
  {title} {\bibinfo {title} {{Tunable strongly coupled superconductivity in
  magic-angle twisted trilayer graphene}},\ }\href
  {https://doi.org/10.1038/s41586-021-03192-0} {\bibfield  {journal} {\bibinfo
  {journal} {\nat}\ }\textbf {\bibinfo {volume} {590}},\ \bibinfo {pages} {249}
  (\bibinfo {year} {2021})}\BibitemShut {NoStop}%
\bibitem [{\citenamefont {{Park}}\ \emph {et~al.}(2022)\citenamefont {{Park}},
  \citenamefont {{Cao}}, \citenamefont {{Xia}}, \citenamefont {{Sun}},
  \citenamefont {{Watanabe}}, \citenamefont {{Taniguchi}},\ and\ \citenamefont
  {{Jarillo-Herrero}}}]{park2022family}%
  \BibitemOpen
  \bibfield  {author} {\bibinfo {author} {\bibfnamefont {J.~M.}\ \bibnamefont
  {{Park}}}, \bibinfo {author} {\bibfnamefont {Y.}~\bibnamefont {{Cao}}},
  \bibinfo {author} {\bibfnamefont {L.-Q.}\ \bibnamefont {{Xia}}}, \bibinfo
  {author} {\bibfnamefont {S.}~\bibnamefont {{Sun}}}, \bibinfo {author}
  {\bibfnamefont {K.}~\bibnamefont {{Watanabe}}}, \bibinfo {author}
  {\bibfnamefont {T.}~\bibnamefont {{Taniguchi}}},\ and\ \bibinfo {author}
  {\bibfnamefont {P.}~\bibnamefont {{Jarillo-Herrero}}},\ }\bibfield  {title}
  {\bibinfo {title} {{Robust superconductivity in magic-angle multilayer
  graphene family}},\ }\href {https://doi.org/10.1038/s41563-022-01287-1}
  {\bibfield  {journal} {\bibinfo  {journal} {Nature Materials}\ }\textbf
  {\bibinfo {volume} {21}},\ \bibinfo {pages} {877} (\bibinfo {year}
  {2022})}\BibitemShut {NoStop}%
\bibitem [{\citenamefont {{Zhang}}\ \emph {et~al.}(2021)\citenamefont
  {{Zhang}}, \citenamefont {{Polski}}, \citenamefont {{Lewandowski}},
  \citenamefont {{Thomson}}, \citenamefont {{Peng}}, \citenamefont {{Choi}},
  \citenamefont {{Kim}}, \citenamefont {{Watanabe}}, \citenamefont
  {{Taniguchi}}, \citenamefont {{Alicea}}, \citenamefont {{von Oppen}},
  \citenamefont {{Refael}},\ and\ \citenamefont
  {{Nadj-Perge}}}]{zhang2021ascendance}%
  \BibitemOpen
  \bibfield  {author} {\bibinfo {author} {\bibfnamefont {Y.}~\bibnamefont
  {{Zhang}}}, \bibinfo {author} {\bibfnamefont {R.}~\bibnamefont {{Polski}}},
  \bibinfo {author} {\bibfnamefont {C.}~\bibnamefont {{Lewandowski}}}, \bibinfo
  {author} {\bibfnamefont {A.}~\bibnamefont {{Thomson}}}, \bibinfo {author}
  {\bibfnamefont {Y.}~\bibnamefont {{Peng}}}, \bibinfo {author} {\bibfnamefont
  {Y.}~\bibnamefont {{Choi}}}, \bibinfo {author} {\bibfnamefont
  {H.}~\bibnamefont {{Kim}}}, \bibinfo {author} {\bibfnamefont
  {K.}~\bibnamefont {{Watanabe}}}, \bibinfo {author} {\bibfnamefont
  {T.}~\bibnamefont {{Taniguchi}}}, \bibinfo {author} {\bibfnamefont
  {J.}~\bibnamefont {{Alicea}}}, \bibinfo {author} {\bibfnamefont
  {F.}~\bibnamefont {{von Oppen}}}, \bibinfo {author} {\bibfnamefont
  {G.}~\bibnamefont {{Refael}}},\ and\ \bibinfo {author} {\bibfnamefont
  {S.}~\bibnamefont {{Nadj-Perge}}},\ }\bibfield  {title} {\bibinfo {title}
  {{Ascendance of Superconductivity in Magic-Angle Graphene Multilayers}},\
  }\href@noop {} {\bibfield  {journal} {\bibinfo  {journal} {arXiv e-prints}\
  ,\ \bibinfo {eid} {arXiv:2112.09270}} (\bibinfo {year} {2021})},\ \Eprint
  {https://arxiv.org/abs/2112.09270} {arXiv:2112.09270 [cond-mat.supr-con]}
  \BibitemShut {NoStop}%
\bibitem [{\citenamefont {Burg}\ \emph {et~al.}(2022)\citenamefont {Burg},
  \citenamefont {Khalaf}, \citenamefont {Wang}, \citenamefont {Watanabe},
  \citenamefont {Taniguchi},\ and\ \citenamefont
  {Tutuc}}]{burg_emergence_2022}%
  \BibitemOpen
  \bibfield  {author} {\bibinfo {author} {\bibfnamefont {G.~W.}\ \bibnamefont
  {Burg}}, \bibinfo {author} {\bibfnamefont {E.}~\bibnamefont {Khalaf}},
  \bibinfo {author} {\bibfnamefont {Y.}~\bibnamefont {Wang}}, \bibinfo {author}
  {\bibfnamefont {K.}~\bibnamefont {Watanabe}}, \bibinfo {author}
  {\bibfnamefont {T.}~\bibnamefont {Taniguchi}},\ and\ \bibinfo {author}
  {\bibfnamefont {E.}~\bibnamefont {Tutuc}},\ }\bibfield  {title} {\bibinfo
  {title} {Emergence of correlations in alternating twist quadrilayer
  graphene},\ }\href {https://doi.org/10.1038/s41563-022-01286-2} {\bibfield
  {journal} {\bibinfo  {journal} {Nature Materials}\ }\textbf {\bibinfo
  {volume} {21}},\ \bibinfo {pages} {884} (\bibinfo {year} {2022})}\BibitemShut
  {NoStop}%
\bibitem [{\citenamefont {{Khalaf}}\ \emph {et~al.}(2019)\citenamefont
  {{Khalaf}}, \citenamefont {{Kruchkov}}, \citenamefont {{Tarnopolsky}},\ and\
  \citenamefont {{Vishwanath}}}]{khalaf2019}%
  \BibitemOpen
  \bibfield  {author} {\bibinfo {author} {\bibfnamefont {E.}~\bibnamefont
  {{Khalaf}}}, \bibinfo {author} {\bibfnamefont {A.~J.}\ \bibnamefont
  {{Kruchkov}}}, \bibinfo {author} {\bibfnamefont {G.}~\bibnamefont
  {{Tarnopolsky}}},\ and\ \bibinfo {author} {\bibfnamefont {A.}~\bibnamefont
  {{Vishwanath}}},\ }\bibfield  {title} {\bibinfo {title} {{Magic angle
  hierarchy in twisted graphene multilayers}},\ }\href
  {https://doi.org/10.1103/PhysRevB.100.085109} {\bibfield  {journal} {\bibinfo
   {journal} {\prb}\ }\textbf {\bibinfo {volume} {100}},\ \bibinfo {eid}
  {085109} (\bibinfo {year} {2019})},\ \Eprint
  {https://arxiv.org/abs/1901.10485} {arXiv:1901.10485 [cond-mat.str-el]}
  \BibitemShut {NoStop}%
\bibitem [{\citenamefont {Zhang}\ \emph {et~al.}(2022)\citenamefont {Zhang},
  \citenamefont {Polski}, \citenamefont {Lewandowski}, \citenamefont {Thomson},
  \citenamefont {Peng}, \citenamefont {Choi}, \citenamefont {Kim},
  \citenamefont {Watanabe}, \citenamefont {Taniguchi}, \citenamefont {Alicea},
  \citenamefont {von Oppen}, \citenamefont {Refael},\ and\ \citenamefont
  {Nadj-Perge}}]{zhang_promotion_2022}%
  \BibitemOpen
  \bibfield  {author} {\bibinfo {author} {\bibfnamefont {Y.}~\bibnamefont
  {Zhang}}, \bibinfo {author} {\bibfnamefont {R.}~\bibnamefont {Polski}},
  \bibinfo {author} {\bibfnamefont {C.}~\bibnamefont {Lewandowski}}, \bibinfo
  {author} {\bibfnamefont {A.}~\bibnamefont {Thomson}}, \bibinfo {author}
  {\bibfnamefont {Y.}~\bibnamefont {Peng}}, \bibinfo {author} {\bibfnamefont
  {Y.}~\bibnamefont {Choi}}, \bibinfo {author} {\bibfnamefont {H.}~\bibnamefont
  {Kim}}, \bibinfo {author} {\bibfnamefont {K.}~\bibnamefont {Watanabe}},
  \bibinfo {author} {\bibfnamefont {T.}~\bibnamefont {Taniguchi}}, \bibinfo
  {author} {\bibfnamefont {J.}~\bibnamefont {Alicea}}, \bibinfo {author}
  {\bibfnamefont {F.}~\bibnamefont {von Oppen}}, \bibinfo {author}
  {\bibfnamefont {G.}~\bibnamefont {Refael}},\ and\ \bibinfo {author}
  {\bibfnamefont {S.}~\bibnamefont {Nadj-Perge}},\ }\bibfield  {title}
  {\bibinfo {title} {Promotion of superconductivity in magic-angle graphene
  multilayers},\ }\href {https://doi.org/10.1126/science.abn8585} {\bibfield
  {journal} {\bibinfo  {journal} {Science}\ }\textbf {\bibinfo {volume}
  {377}},\ \bibinfo {pages} {1538} (\bibinfo {year} {2022})}\BibitemShut
  {NoStop}%
\bibitem [{\citenamefont {{Kim}}\ \emph {et~al.}(2022)\citenamefont {{Kim}},
  \citenamefont {{Choi}}, \citenamefont {{Lewandowski}}, \citenamefont
  {{Thomson}}, \citenamefont {{Zhang}}, \citenamefont {{Polski}}, \citenamefont
  {{Watanabe}}, \citenamefont {{Taniguchi}}, \citenamefont {{Alicea}},\ and\
  \citenamefont {{Nadj-Perge}}}]{kim2022}%
  \BibitemOpen
  \bibfield  {author} {\bibinfo {author} {\bibfnamefont {H.}~\bibnamefont
  {{Kim}}}, \bibinfo {author} {\bibfnamefont {Y.}~\bibnamefont {{Choi}}},
  \bibinfo {author} {\bibfnamefont {C.}~\bibnamefont {{Lewandowski}}}, \bibinfo
  {author} {\bibfnamefont {A.}~\bibnamefont {{Thomson}}}, \bibinfo {author}
  {\bibfnamefont {Y.}~\bibnamefont {{Zhang}}}, \bibinfo {author} {\bibfnamefont
  {R.}~\bibnamefont {{Polski}}}, \bibinfo {author} {\bibfnamefont
  {K.}~\bibnamefont {{Watanabe}}}, \bibinfo {author} {\bibfnamefont
  {T.}~\bibnamefont {{Taniguchi}}}, \bibinfo {author} {\bibfnamefont
  {J.}~\bibnamefont {{Alicea}}},\ and\ \bibinfo {author} {\bibfnamefont
  {S.}~\bibnamefont {{Nadj-Perge}}},\ }\bibfield  {title} {\bibinfo {title}
  {{Evidence for unconventional superconductivity in twisted trilayer
  graphene}},\ }\href {https://doi.org/10.1038/s41586-022-04715-z} {\bibfield
  {journal} {\bibinfo  {journal} {\nat}\ }\textbf {\bibinfo {volume} {606}},\
  \bibinfo {pages} {494} (\bibinfo {year} {2022})}\BibitemShut {NoStop}%
\bibitem [{\citenamefont {{Rozen}}\ \emph {et~al.}(2021)\citenamefont
  {{Rozen}}, \citenamefont {{Park}}, \citenamefont {{Zondiner}}, \citenamefont
  {{Cao}}, \citenamefont {{Rodan-Legrain}}, \citenamefont {{Taniguchi}},
  \citenamefont {{Watanabe}}, \citenamefont {{Oreg}}, \citenamefont {{Stern}},
  \citenamefont {{Berg}}, \citenamefont {{Jarillo-Herrero}},\ and\
  \citenamefont {{Ilani}}}]{rozen2021}%
  \BibitemOpen
  \bibfield  {author} {\bibinfo {author} {\bibfnamefont {A.}~\bibnamefont
  {{Rozen}}}, \bibinfo {author} {\bibfnamefont {J.~M.}\ \bibnamefont {{Park}}},
  \bibinfo {author} {\bibfnamefont {U.}~\bibnamefont {{Zondiner}}}, \bibinfo
  {author} {\bibfnamefont {Y.}~\bibnamefont {{Cao}}}, \bibinfo {author}
  {\bibfnamefont {D.}~\bibnamefont {{Rodan-Legrain}}}, \bibinfo {author}
  {\bibfnamefont {T.}~\bibnamefont {{Taniguchi}}}, \bibinfo {author}
  {\bibfnamefont {K.}~\bibnamefont {{Watanabe}}}, \bibinfo {author}
  {\bibfnamefont {Y.}~\bibnamefont {{Oreg}}}, \bibinfo {author} {\bibfnamefont
  {A.}~\bibnamefont {{Stern}}}, \bibinfo {author} {\bibfnamefont
  {E.}~\bibnamefont {{Berg}}}, \bibinfo {author} {\bibfnamefont
  {P.}~\bibnamefont {{Jarillo-Herrero}}},\ and\ \bibinfo {author}
  {\bibfnamefont {S.}~\bibnamefont {{Ilani}}},\ }\bibfield  {title} {\bibinfo
  {title} {{Entropic evidence for a Pomeranchuk effect in magic-angle
  graphene}},\ }\href {https://doi.org/10.1038/s41586-021-03319-3} {\bibfield
  {journal} {\bibinfo  {journal} {\nat}\ }\textbf {\bibinfo {volume} {592}},\
  \bibinfo {pages} {214} (\bibinfo {year} {2021})},\ \Eprint
  {https://arxiv.org/abs/2009.01836} {arXiv:2009.01836 [cond-mat.mes-hall]}
  \BibitemShut {NoStop}%
\bibitem [{\citenamefont {{Kerelsky}}\ \emph {et~al.}(2019)\citenamefont
  {{Kerelsky}}, \citenamefont {{McGilly}}, \citenamefont {{Kennes}},
  \citenamefont {{Xian}}, \citenamefont {{Yankowitz}}, \citenamefont
  {{Shaowen}}, \citenamefont {{Watanabe}}, \citenamefont {{Taniguchi}},
  \citenamefont {{Hone}}, \citenamefont {{Rubio}}, \citenamefont {{Dean}},\
  and\ \citenamefont {{Pasupathy}}}]{kerelsky2019}%
  \BibitemOpen
  \bibfield  {author} {\bibinfo {author} {\bibfnamefont {A.}~\bibnamefont
  {{Kerelsky}}}, \bibinfo {author} {\bibfnamefont {L.}~\bibnamefont
  {{McGilly}}}, \bibinfo {author} {\bibfnamefont {D.}~\bibnamefont {{Kennes}}},
  \bibinfo {author} {\bibfnamefont {L.}~\bibnamefont {{Xian}}}, \bibinfo
  {author} {\bibfnamefont {M.}~\bibnamefont {{Yankowitz}}}, \bibinfo {author}
  {\bibfnamefont {C.}~\bibnamefont {{Shaowen}}}, \bibinfo {author}
  {\bibfnamefont {K.}~\bibnamefont {{Watanabe}}}, \bibinfo {author}
  {\bibfnamefont {T.}~\bibnamefont {{Taniguchi}}}, \bibinfo {author}
  {\bibfnamefont {J.}~\bibnamefont {{Hone}}}, \bibinfo {author} {\bibfnamefont
  {A.}~\bibnamefont {{Rubio}}}, \bibinfo {author} {\bibfnamefont
  {C.}~\bibnamefont {{Dean}}},\ and\ \bibinfo {author} {\bibfnamefont
  {A.}~\bibnamefont {{Pasupathy}}},\ }\bibfield  {title} {\bibinfo {title}
  {{Maximized electron interactions at the magic angle in twisted bilayer
  graphene}},\ }\href@noop {} {\bibfield  {journal} {\bibinfo  {journal}
  {\nat}\ }\textbf {\bibinfo {volume} {572}},\ \bibinfo {pages} {95} (\bibinfo
  {year} {2019})}\BibitemShut {NoStop}%
\bibitem [{\citenamefont {{Xie}}\ \emph {et~al.}(2019)\citenamefont {{Xie}},
  \citenamefont {{Lian}}, \citenamefont {{J{\"a}ck}}, \citenamefont {{Liu}},
  \citenamefont {{Chiu}}, \citenamefont {{Watanabe}}, \citenamefont
  {{Taniguchi}}, \citenamefont {{Bernevig}},\ and\ \citenamefont
  {{Yazdani}}}]{xie2019}%
  \BibitemOpen
  \bibfield  {author} {\bibinfo {author} {\bibfnamefont {Y.}~\bibnamefont
  {{Xie}}}, \bibinfo {author} {\bibfnamefont {B.}~\bibnamefont {{Lian}}},
  \bibinfo {author} {\bibfnamefont {B.}~\bibnamefont {{J{\"a}ck}}}, \bibinfo
  {author} {\bibfnamefont {X.}~\bibnamefont {{Liu}}}, \bibinfo {author}
  {\bibfnamefont {C.-L.}\ \bibnamefont {{Chiu}}}, \bibinfo {author}
  {\bibfnamefont {K.}~\bibnamefont {{Watanabe}}}, \bibinfo {author}
  {\bibfnamefont {T.}~\bibnamefont {{Taniguchi}}}, \bibinfo {author}
  {\bibfnamefont {B.~A.}\ \bibnamefont {{Bernevig}}},\ and\ \bibinfo {author}
  {\bibfnamefont {A.}~\bibnamefont {{Yazdani}}},\ }\bibfield  {title} {\bibinfo
  {title} {{Spectroscopic signatures of many-body correlations in magic-angle
  twisted bilayer graphene}},\ }\href
  {https://doi.org/10.1038/s41586-019-1422-x} {\bibfield  {journal} {\bibinfo
  {journal} {\nat}\ }\textbf {\bibinfo {volume} {572}},\ \bibinfo {pages} {101}
  (\bibinfo {year} {2019})},\ \Eprint {https://arxiv.org/abs/1906.09274}
  {arXiv:1906.09274 [cond-mat.mes-hall]} \BibitemShut {NoStop}%
\bibitem [{\citenamefont {{Jiang}}\ \emph {et~al.}(2019)\citenamefont
  {{Jiang}}, \citenamefont {{Lai}}, \citenamefont {{Watanabe}}, \citenamefont
  {{Taniguchi}}, \citenamefont {{Haule}}, \citenamefont {{Mao}},\ and\
  \citenamefont {{Andrei}}}]{jiang2019}%
  \BibitemOpen
  \bibfield  {author} {\bibinfo {author} {\bibfnamefont {Y.}~\bibnamefont
  {{Jiang}}}, \bibinfo {author} {\bibfnamefont {X.}~\bibnamefont {{Lai}}},
  \bibinfo {author} {\bibfnamefont {K.}~\bibnamefont {{Watanabe}}}, \bibinfo
  {author} {\bibfnamefont {T.}~\bibnamefont {{Taniguchi}}}, \bibinfo {author}
  {\bibfnamefont {K.}~\bibnamefont {{Haule}}}, \bibinfo {author} {\bibfnamefont
  {J.}~\bibnamefont {{Mao}}},\ and\ \bibinfo {author} {\bibfnamefont {E.~Y.}\
  \bibnamefont {{Andrei}}},\ }\bibfield  {title} {\bibinfo {title} {{Charge
  order and broken rotational symmetry in magic-angle twisted bilayer
  graphene}},\ }\href {https://doi.org/10.1038/s41586-019-1460-4} {\bibfield
  {journal} {\bibinfo  {journal} {\nat}\ }\textbf {\bibinfo {volume} {573}},\
  \bibinfo {pages} {91} (\bibinfo {year} {2019})},\ \Eprint
  {https://arxiv.org/abs/1904.10153} {arXiv:1904.10153 [cond-mat.mes-hall]}
  \BibitemShut {NoStop}%
\bibitem [{\citenamefont {{Choi}}\ \emph {et~al.}(2019)\citenamefont {{Choi}},
  \citenamefont {{Kemmer}}, \citenamefont {{Peng}}, \citenamefont {{Thomson}},
  \citenamefont {{Arora}}, \citenamefont {{Polski}}, \citenamefont {{Zhang}},
  \citenamefont {{Ren}}, \citenamefont {{Alicea}}, \citenamefont {{Refael}},
  \citenamefont {{von Oppen}}, \citenamefont {{Watanabe}}, \citenamefont
  {{Taniguchi}},\ and\ \citenamefont {{Nadj-Perge}}}]{choi2019stm}%
  \BibitemOpen
  \bibfield  {author} {\bibinfo {author} {\bibfnamefont {Y.}~\bibnamefont
  {{Choi}}}, \bibinfo {author} {\bibfnamefont {J.}~\bibnamefont {{Kemmer}}},
  \bibinfo {author} {\bibfnamefont {Y.}~\bibnamefont {{Peng}}}, \bibinfo
  {author} {\bibfnamefont {A.}~\bibnamefont {{Thomson}}}, \bibinfo {author}
  {\bibfnamefont {H.}~\bibnamefont {{Arora}}}, \bibinfo {author} {\bibfnamefont
  {R.}~\bibnamefont {{Polski}}}, \bibinfo {author} {\bibfnamefont
  {Y.}~\bibnamefont {{Zhang}}}, \bibinfo {author} {\bibfnamefont
  {H.}~\bibnamefont {{Ren}}}, \bibinfo {author} {\bibfnamefont
  {J.}~\bibnamefont {{Alicea}}}, \bibinfo {author} {\bibfnamefont
  {G.}~\bibnamefont {{Refael}}}, \bibinfo {author} {\bibfnamefont
  {F.}~\bibnamefont {{von Oppen}}}, \bibinfo {author} {\bibfnamefont
  {K.}~\bibnamefont {{Watanabe}}}, \bibinfo {author} {\bibfnamefont
  {T.}~\bibnamefont {{Taniguchi}}},\ and\ \bibinfo {author} {\bibfnamefont
  {S.}~\bibnamefont {{Nadj-Perge}}},\ }\bibfield  {title} {\bibinfo {title}
  {{Electronic correlations in twisted bilayer graphene near the magic
  angle}},\ }\href {https://doi.org/10.1038/s41567-019-0606-5} {\bibfield
  {journal} {\bibinfo  {journal} {Nature Physics}\ }\textbf {\bibinfo {volume}
  {15}},\ \bibinfo {pages} {1174} (\bibinfo {year} {2019})},\ \Eprint
  {https://arxiv.org/abs/1901.02997} {arXiv:1901.02997 [cond-mat.mes-hall]}
  \BibitemShut {NoStop}%
\bibitem [{\citenamefont {{Nuckolls}}\ \emph {et~al.}(2023)\citenamefont
  {{Nuckolls}}, \citenamefont {{Lee}}, \citenamefont {{Oh}}, \citenamefont
  {{Wong}}, \citenamefont {{Soejima}}, \citenamefont {{Hong}}, \citenamefont
  {{C{\v{a}}lug{\v{a}}ru}}, \citenamefont {{Herzog-Arbeitman}}, \citenamefont
  {{Bernevig}}, \citenamefont {{Watanabe}}, \citenamefont {{Taniguchi}},
  \citenamefont {{Regnault}}, \citenamefont {{Zaletel}},\ and\ \citenamefont
  {{Yazdani}}}]{nuckolls2023}%
  \BibitemOpen
  \bibfield  {author} {\bibinfo {author} {\bibfnamefont {K.~P.}\ \bibnamefont
  {{Nuckolls}}}, \bibinfo {author} {\bibfnamefont {R.~L.}\ \bibnamefont
  {{Lee}}}, \bibinfo {author} {\bibfnamefont {M.}~\bibnamefont {{Oh}}},
  \bibinfo {author} {\bibfnamefont {D.}~\bibnamefont {{Wong}}}, \bibinfo
  {author} {\bibfnamefont {T.}~\bibnamefont {{Soejima}}}, \bibinfo {author}
  {\bibfnamefont {J.~P.}\ \bibnamefont {{Hong}}}, \bibinfo {author}
  {\bibfnamefont {D.}~\bibnamefont {{C{\v{a}}lug{\v{a}}ru}}}, \bibinfo {author}
  {\bibfnamefont {J.}~\bibnamefont {{Herzog-Arbeitman}}}, \bibinfo {author}
  {\bibfnamefont {B.~A.}\ \bibnamefont {{Bernevig}}}, \bibinfo {author}
  {\bibfnamefont {K.}~\bibnamefont {{Watanabe}}}, \bibinfo {author}
  {\bibfnamefont {T.}~\bibnamefont {{Taniguchi}}}, \bibinfo {author}
  {\bibfnamefont {N.}~\bibnamefont {{Regnault}}}, \bibinfo {author}
  {\bibfnamefont {M.~P.}\ \bibnamefont {{Zaletel}}},\ and\ \bibinfo {author}
  {\bibfnamefont {A.}~\bibnamefont {{Yazdani}}},\ }\bibfield  {title} {\bibinfo
  {title} {{Quantum textures of the many-body wavefunctions in magic-angle
  graphene}},\ }\href {https://doi.org/10.1038/s41586-023-06226-x} {\bibfield
  {journal} {\bibinfo  {journal} {\nat}\ }\textbf {\bibinfo {volume} {620}},\
  \bibinfo {pages} {525} (\bibinfo {year} {2023})},\ \Eprint
  {https://arxiv.org/abs/2303.00024} {arXiv:2303.00024 [cond-mat.mes-hall]}
  \BibitemShut {NoStop}%
\bibitem [{\citenamefont {{Polshyn}}\ \emph {et~al.}(2019)\citenamefont
  {{Polshyn}}, \citenamefont {{Yankowitz}}, \citenamefont {{Chen}},
  \citenamefont {{Zhang}}, \citenamefont {{Watanabe}}, \citenamefont
  {{Taniguchi}}, \citenamefont {{Dean}},\ and\ \citenamefont
  {{Young}}}]{polshyn2019}%
  \BibitemOpen
  \bibfield  {author} {\bibinfo {author} {\bibfnamefont {H.}~\bibnamefont
  {{Polshyn}}}, \bibinfo {author} {\bibfnamefont {M.}~\bibnamefont
  {{Yankowitz}}}, \bibinfo {author} {\bibfnamefont {S.}~\bibnamefont {{Chen}}},
  \bibinfo {author} {\bibfnamefont {Y.}~\bibnamefont {{Zhang}}}, \bibinfo
  {author} {\bibfnamefont {K.}~\bibnamefont {{Watanabe}}}, \bibinfo {author}
  {\bibfnamefont {T.}~\bibnamefont {{Taniguchi}}}, \bibinfo {author}
  {\bibfnamefont {C.~R.}\ \bibnamefont {{Dean}}},\ and\ \bibinfo {author}
  {\bibfnamefont {A.~F.}\ \bibnamefont {{Young}}},\ }\bibfield  {title}
  {\bibinfo {title} {{Large linear-in-temperature resistivity in twisted
  bilayer graphene}},\ }\href {https://doi.org/10.1038/s41567-019-0596-3}
  {\bibfield  {journal} {\bibinfo  {journal} {Nature Physics}\ }\textbf
  {\bibinfo {volume} {15}},\ \bibinfo {pages} {1011} (\bibinfo {year}
  {2019})}\BibitemShut {NoStop}%
\bibitem [{\citenamefont {Wu}\ \emph {et~al.}(2018)\citenamefont {Wu},
  \citenamefont {MacDonald},\ and\ \citenamefont {Martin}}]{AllanPhononDriven}%
  \BibitemOpen
  \bibfield  {author} {\bibinfo {author} {\bibfnamefont {F.}~\bibnamefont
  {Wu}}, \bibinfo {author} {\bibfnamefont {A.~H.}\ \bibnamefont {MacDonald}},\
  and\ \bibinfo {author} {\bibfnamefont {I.}~\bibnamefont {Martin}},\
  }\bibfield  {title} {\bibinfo {title} {Theory of phonon-mediated
  superconductivity in twisted bilayer graphene},\ }\href
  {https://doi.org/10.1103/PhysRevLett.121.257001} {\bibfield  {journal}
  {\bibinfo  {journal} {Phys. Rev. Lett.}\ }\textbf {\bibinfo {volume} {121}},\
  \bibinfo {pages} {257001} (\bibinfo {year} {2018})}\BibitemShut {NoStop}%
\bibitem [{\citenamefont {{Wu}}\ \emph {et~al.}(2019)\citenamefont {{Wu}},
  \citenamefont {{Hwang}},\ and\ \citenamefont {{Das Sarma}}}]{WuPhonondriven}%
  \BibitemOpen
  \bibfield  {author} {\bibinfo {author} {\bibfnamefont {F.}~\bibnamefont
  {{Wu}}}, \bibinfo {author} {\bibfnamefont {E.}~\bibnamefont {{Hwang}}},\ and\
  \bibinfo {author} {\bibfnamefont {S.}~\bibnamefont {{Das Sarma}}},\
  }\bibfield  {title} {\bibinfo {title} {{Phonon-induced giant linear-in-T
  resistivity in magic angle twisted bilayer graphene: Ordinary strangeness and
  exotic superconductivity}},\ }\href
  {https://doi.org/10.1103/PhysRevB.99.165112} {\bibfield  {journal} {\bibinfo
  {journal} {\prb}\ }\textbf {\bibinfo {volume} {99}},\ \bibinfo {eid} {165112}
  (\bibinfo {year} {2019})},\ \Eprint {https://arxiv.org/abs/1811.04920}
  {arXiv:1811.04920 [cond-mat.mes-hall]} \BibitemShut {NoStop}%
\bibitem [{\citenamefont {{Chen}}\ \emph {et~al.}(2024)\citenamefont {{Chen}},
  \citenamefont {{Nuckolls}}, \citenamefont {{Ding}}, \citenamefont {{Miao}},
  \citenamefont {{Wong}}, \citenamefont {{Oh}}, \citenamefont {{Lee}},
  \citenamefont {{He}}, \citenamefont {{Peng}}, \citenamefont {{Pei}},
  \citenamefont {{Li}}, \citenamefont {{Hao}}, \citenamefont {{Yan}},
  \citenamefont {{Xiao}}, \citenamefont {{Gao}}, \citenamefont {{Li}},
  \citenamefont {{Zhang}}, \citenamefont {{Liu}}, \citenamefont {{He}},
  \citenamefont {{Watanabe}}, \citenamefont {{Taniguchi}}, \citenamefont
  {{Jozwiak}}, \citenamefont {{Bostwick}}, \citenamefont {{Rotenberg}},
  \citenamefont {{Li}}, \citenamefont {{Han}}, \citenamefont {{Pan}},
  \citenamefont {{Liu}}, \citenamefont {{Dai}}, \citenamefont {{Liu}},
  \citenamefont {{Bernevig}}, \citenamefont {{Wang}}, \citenamefont
  {{Yazdani}},\ and\ \citenamefont {{Chen}}}]{Chen2024phonon}%
  \BibitemOpen
  \bibfield  {author} {\bibinfo {author} {\bibfnamefont {C.}~\bibnamefont
  {{Chen}}}, \bibinfo {author} {\bibfnamefont {K.~P.}\ \bibnamefont
  {{Nuckolls}}}, \bibinfo {author} {\bibfnamefont {S.}~\bibnamefont {{Ding}}},
  \bibinfo {author} {\bibfnamefont {W.}~\bibnamefont {{Miao}}}, \bibinfo
  {author} {\bibfnamefont {D.}~\bibnamefont {{Wong}}}, \bibinfo {author}
  {\bibfnamefont {M.}~\bibnamefont {{Oh}}}, \bibinfo {author} {\bibfnamefont
  {R.~L.}\ \bibnamefont {{Lee}}}, \bibinfo {author} {\bibfnamefont
  {S.}~\bibnamefont {{He}}}, \bibinfo {author} {\bibfnamefont {C.}~\bibnamefont
  {{Peng}}}, \bibinfo {author} {\bibfnamefont {D.}~\bibnamefont {{Pei}}},
  \bibinfo {author} {\bibfnamefont {Y.}~\bibnamefont {{Li}}}, \bibinfo {author}
  {\bibfnamefont {C.}~\bibnamefont {{Hao}}}, \bibinfo {author} {\bibfnamefont
  {H.}~\bibnamefont {{Yan}}}, \bibinfo {author} {\bibfnamefont
  {H.}~\bibnamefont {{Xiao}}}, \bibinfo {author} {\bibfnamefont
  {H.}~\bibnamefont {{Gao}}}, \bibinfo {author} {\bibfnamefont
  {Q.}~\bibnamefont {{Li}}}, \bibinfo {author} {\bibfnamefont {S.}~\bibnamefont
  {{Zhang}}}, \bibinfo {author} {\bibfnamefont {J.}~\bibnamefont {{Liu}}},
  \bibinfo {author} {\bibfnamefont {L.}~\bibnamefont {{He}}}, \bibinfo {author}
  {\bibfnamefont {K.}~\bibnamefont {{Watanabe}}}, \bibinfo {author}
  {\bibfnamefont {T.}~\bibnamefont {{Taniguchi}}}, \bibinfo {author}
  {\bibfnamefont {C.}~\bibnamefont {{Jozwiak}}}, \bibinfo {author}
  {\bibfnamefont {A.}~\bibnamefont {{Bostwick}}}, \bibinfo {author}
  {\bibfnamefont {E.}~\bibnamefont {{Rotenberg}}}, \bibinfo {author}
  {\bibfnamefont {C.}~\bibnamefont {{Li}}}, \bibinfo {author} {\bibfnamefont
  {X.}~\bibnamefont {{Han}}}, \bibinfo {author} {\bibfnamefont
  {D.}~\bibnamefont {{Pan}}}, \bibinfo {author} {\bibfnamefont
  {Z.}~\bibnamefont {{Liu}}}, \bibinfo {author} {\bibfnamefont
  {X.}~\bibnamefont {{Dai}}}, \bibinfo {author} {\bibfnamefont
  {C.}~\bibnamefont {{Liu}}}, \bibinfo {author} {\bibfnamefont {B.~A.}\
  \bibnamefont {{Bernevig}}}, \bibinfo {author} {\bibfnamefont
  {Y.}~\bibnamefont {{Wang}}}, \bibinfo {author} {\bibfnamefont
  {A.}~\bibnamefont {{Yazdani}}},\ and\ \bibinfo {author} {\bibfnamefont
  {Y.}~\bibnamefont {{Chen}}},\ }\bibfield  {title} {\bibinfo {title} {{Strong
  electron{\textendash}phonon coupling in magic-angle twisted bilayer
  graphene}},\ }\href {https://doi.org/10.1038/s41586-024-08227-w} {\bibfield
  {journal} {\bibinfo  {journal} {\nat}\ }\textbf {\bibinfo {volume} {636}},\
  \bibinfo {pages} {342} (\bibinfo {year} {2024})},\ \Eprint
  {https://arxiv.org/abs/2303.14903} {arXiv:2303.14903 [cond-mat.mes-hall]}
  \BibitemShut {NoStop}%
\bibitem [{\citenamefont {{Kwan}}\ \emph {et~al.}(2024)\citenamefont {{Kwan}},
  \citenamefont {{Wagner}}, \citenamefont {{Bultinck}}, \citenamefont
  {{Simon}}, \citenamefont {{Berg}},\ and\ \citenamefont
  {{Parameswaran}}}]{kwan2024kekulephonon}%
  \BibitemOpen
  \bibfield  {author} {\bibinfo {author} {\bibfnamefont {Y.~H.}\ \bibnamefont
  {{Kwan}}}, \bibinfo {author} {\bibfnamefont {G.}~\bibnamefont {{Wagner}}},
  \bibinfo {author} {\bibfnamefont {N.}~\bibnamefont {{Bultinck}}}, \bibinfo
  {author} {\bibfnamefont {S.~H.}\ \bibnamefont {{Simon}}}, \bibinfo {author}
  {\bibfnamefont {E.}~\bibnamefont {{Berg}}},\ and\ \bibinfo {author}
  {\bibfnamefont {S.~A.}\ \bibnamefont {{Parameswaran}}},\ }\bibfield  {title}
  {\bibinfo {title} {{Electron-phonon coupling and competing Kekul{\'e} orders
  in twisted bilayer graphene}},\ }\href
  {https://doi.org/10.1103/PhysRevB.110.085160} {\bibfield  {journal} {\bibinfo
   {journal} {\prb}\ }\textbf {\bibinfo {volume} {110}},\ \bibinfo {eid}
  {085160} (\bibinfo {year} {2024})},\ \Eprint
  {https://arxiv.org/abs/2303.13602} {arXiv:2303.13602 [cond-mat.str-el]}
  \BibitemShut {NoStop}%
\bibitem [{\citenamefont {{Ledwith}}\ \emph {et~al.}(2021)\citenamefont
  {{Ledwith}}, \citenamefont {{Khalaf}}, \citenamefont {{Zhu}}, \citenamefont
  {{Carr}}, \citenamefont {{Kaxiras}},\ and\ \citenamefont
  {{Vishwanath}}}]{ledwith2021tb}%
  \BibitemOpen
  \bibfield  {author} {\bibinfo {author} {\bibfnamefont {P.~J.}\ \bibnamefont
  {{Ledwith}}}, \bibinfo {author} {\bibfnamefont {E.}~\bibnamefont {{Khalaf}}},
  \bibinfo {author} {\bibfnamefont {Z.}~\bibnamefont {{Zhu}}}, \bibinfo
  {author} {\bibfnamefont {S.}~\bibnamefont {{Carr}}}, \bibinfo {author}
  {\bibfnamefont {E.}~\bibnamefont {{Kaxiras}}},\ and\ \bibinfo {author}
  {\bibfnamefont {A.}~\bibnamefont {{Vishwanath}}},\ }\bibfield  {title}
  {\bibinfo {title} {{TB or not TB? Contrasting properties of twisted bilayer
  graphene and the alternating twist $n$-layer structures ($n=3, 4, 5,
  \dots$)}},\ }\href@noop {} {\bibfield  {journal} {\bibinfo  {journal} {arXiv
  e-prints}\ ,\ \bibinfo {eid} {arXiv:2111.11060}} (\bibinfo {year} {2021})},\
  \Eprint {https://arxiv.org/abs/2111.11060} {arXiv:2111.11060
  [cond-mat.str-el]} \BibitemShut {NoStop}%
\bibitem [{\citenamefont {{Zhang}}\ \emph {et~al.}(2019)\citenamefont
  {{Zhang}}, \citenamefont {{Po}},\ and\ \citenamefont
  {{Senthil}}}]{zhang2019_lldegen}%
  \BibitemOpen
  \bibfield  {author} {\bibinfo {author} {\bibfnamefont {Y.-H.}\ \bibnamefont
  {{Zhang}}}, \bibinfo {author} {\bibfnamefont {H.~C.}\ \bibnamefont {{Po}}},\
  and\ \bibinfo {author} {\bibfnamefont {T.}~\bibnamefont {{Senthil}}},\
  }\bibfield  {title} {\bibinfo {title} {{Landau level degeneracy in twisted
  bilayer graphene: Role of symmetry breaking}},\ }\href
  {https://doi.org/10.1103/PhysRevB.100.125104} {\bibfield  {journal} {\bibinfo
   {journal} {\prb}\ }\textbf {\bibinfo {volume} {100}},\ \bibinfo {eid}
  {125104} (\bibinfo {year} {2019})},\ \Eprint
  {https://arxiv.org/abs/1904.10452} {arXiv:1904.10452 [cond-mat.str-el]}
  \BibitemShut {NoStop}%
\bibitem [{\citenamefont {{Zondiner}}\ \emph {et~al.}(2020)\citenamefont
  {{Zondiner}}, \citenamefont {{Rozen}}, \citenamefont {{Rodan-Legrain}},
  \citenamefont {{Cao}}, \citenamefont {{Queiroz}}, \citenamefont
  {{Taniguchi}}, \citenamefont {{Watanabe}}, \citenamefont {{Oreg}},
  \citenamefont {{von Oppen}}, \citenamefont {{Stern}}, \citenamefont {{Berg}},
  \citenamefont {{Jarillo-Herrero}},\ and\ \citenamefont
  {{Ilani}}}]{zondiner2020_cascade}%
  \BibitemOpen
  \bibfield  {author} {\bibinfo {author} {\bibfnamefont {U.}~\bibnamefont
  {{Zondiner}}}, \bibinfo {author} {\bibfnamefont {A.}~\bibnamefont {{Rozen}}},
  \bibinfo {author} {\bibfnamefont {D.}~\bibnamefont {{Rodan-Legrain}}},
  \bibinfo {author} {\bibfnamefont {Y.}~\bibnamefont {{Cao}}}, \bibinfo
  {author} {\bibfnamefont {R.}~\bibnamefont {{Queiroz}}}, \bibinfo {author}
  {\bibfnamefont {T.}~\bibnamefont {{Taniguchi}}}, \bibinfo {author}
  {\bibfnamefont {K.}~\bibnamefont {{Watanabe}}}, \bibinfo {author}
  {\bibfnamefont {Y.}~\bibnamefont {{Oreg}}}, \bibinfo {author} {\bibfnamefont
  {F.}~\bibnamefont {{von Oppen}}}, \bibinfo {author} {\bibfnamefont
  {A.}~\bibnamefont {{Stern}}}, \bibinfo {author} {\bibfnamefont
  {E.}~\bibnamefont {{Berg}}}, \bibinfo {author} {\bibfnamefont
  {P.}~\bibnamefont {{Jarillo-Herrero}}},\ and\ \bibinfo {author}
  {\bibfnamefont {S.}~\bibnamefont {{Ilani}}},\ }\bibfield  {title} {\bibinfo
  {title} {{Cascade of phase transitions and Dirac revivals in magic-angle
  graphene}},\ }\href {https://doi.org/10.1038/s41586-020-2373-y} {\bibfield
  {journal} {\bibinfo  {journal} {\nat}\ }\textbf {\bibinfo {volume} {582}},\
  \bibinfo {pages} {203} (\bibinfo {year} {2020})},\ \Eprint
  {https://arxiv.org/abs/1912.06150} {arXiv:1912.06150 [cond-mat.mes-hall]}
  \BibitemShut {NoStop}%
\bibitem [{\citenamefont {{Park}}\ \emph {et~al.}(2020)\citenamefont {{Park}},
  \citenamefont {{Cao}}, \citenamefont {{Watanabe}}, \citenamefont
  {{Taniguchi}},\ and\ \citenamefont {{Jarillo-Herrero}}}]{park2020hund}%
  \BibitemOpen
  \bibfield  {author} {\bibinfo {author} {\bibfnamefont {J.~M.}\ \bibnamefont
  {{Park}}}, \bibinfo {author} {\bibfnamefont {Y.}~\bibnamefont {{Cao}}},
  \bibinfo {author} {\bibfnamefont {K.}~\bibnamefont {{Watanabe}}}, \bibinfo
  {author} {\bibfnamefont {T.}~\bibnamefont {{Taniguchi}}},\ and\ \bibinfo
  {author} {\bibfnamefont {P.}~\bibnamefont {{Jarillo-Herrero}}},\ }\bibfield
  {title} {\bibinfo {title} {{Flavour Hund's Coupling, Correlated Chern Gaps,
  and Diffusivity in Moir{\'e} Flat Bands}},\ }\href@noop {} {\bibfield
  {journal} {\bibinfo  {journal} {arXiv e-prints}\ ,\ \bibinfo {eid}
  {arXiv:2008.12296}} (\bibinfo {year} {2020})},\ \Eprint
  {https://arxiv.org/abs/2008.12296} {arXiv:2008.12296 [cond-mat.mes-hall]}
  \BibitemShut {NoStop}%
\bibitem [{\citenamefont {{Saito}}\ \emph {et~al.}(2021)\citenamefont
  {{Saito}}, \citenamefont {{Yang}}, \citenamefont {{Ge}}, \citenamefont
  {{Liu}}, \citenamefont {{Taniguchi}}, \citenamefont {{Watanabe}},
  \citenamefont {{Li}}, \citenamefont {{Berg}},\ and\ \citenamefont
  {{Young}}}]{saito2021isospin}%
  \BibitemOpen
  \bibfield  {author} {\bibinfo {author} {\bibfnamefont {Y.}~\bibnamefont
  {{Saito}}}, \bibinfo {author} {\bibfnamefont {F.}~\bibnamefont {{Yang}}},
  \bibinfo {author} {\bibfnamefont {J.}~\bibnamefont {{Ge}}}, \bibinfo {author}
  {\bibfnamefont {X.}~\bibnamefont {{Liu}}}, \bibinfo {author} {\bibfnamefont
  {T.}~\bibnamefont {{Taniguchi}}}, \bibinfo {author} {\bibfnamefont
  {K.}~\bibnamefont {{Watanabe}}}, \bibinfo {author} {\bibfnamefont {J.~I.~A.}\
  \bibnamefont {{Li}}}, \bibinfo {author} {\bibfnamefont {E.}~\bibnamefont
  {{Berg}}},\ and\ \bibinfo {author} {\bibfnamefont {A.~F.}\ \bibnamefont
  {{Young}}},\ }\bibfield  {title} {\bibinfo {title} {{Isospin Pomeranchuk
  effect in twisted bilayer graphene}},\ }\href
  {https://doi.org/10.1038/s41586-021-03409-2} {\bibfield  {journal} {\bibinfo
  {journal} {\nat}\ }\textbf {\bibinfo {volume} {592}},\ \bibinfo {pages} {220}
  (\bibinfo {year} {2021})},\ \Eprint {https://arxiv.org/abs/2008.10830}
  {arXiv:2008.10830 [cond-mat.mes-hall]} \BibitemShut {NoStop}%
\bibitem [{\citenamefont {{Carr}}\ \emph {et~al.}(2020)\citenamefont {{Carr}},
  \citenamefont {{Li}}, \citenamefont {{Zhu}}, \citenamefont {{Kaxiras}},
  \citenamefont {{Sachdev}},\ and\ \citenamefont {{Kruchkov}}}]{carr2020}%
  \BibitemOpen
  \bibfield  {author} {\bibinfo {author} {\bibfnamefont {S.}~\bibnamefont
  {{Carr}}}, \bibinfo {author} {\bibfnamefont {C.}~\bibnamefont {{Li}}},
  \bibinfo {author} {\bibfnamefont {Z.}~\bibnamefont {{Zhu}}}, \bibinfo
  {author} {\bibfnamefont {E.}~\bibnamefont {{Kaxiras}}}, \bibinfo {author}
  {\bibfnamefont {S.}~\bibnamefont {{Sachdev}}},\ and\ \bibinfo {author}
  {\bibfnamefont {A.}~\bibnamefont {{Kruchkov}}},\ }\bibfield  {title}
  {\bibinfo {title} {{Ultraheavy and Ultrarelativistic Dirac Quasiparticles in
  Sandwiched Graphenes}},\ }\href
  {https://doi.org/10.1021/acs.nanolett.9b04979} {\bibfield  {journal}
  {\bibinfo  {journal} {Nano Letters}\ }\textbf {\bibinfo {volume} {20}},\
  \bibinfo {pages} {3030} (\bibinfo {year} {2020})},\ \Eprint
  {https://arxiv.org/abs/1907.00952} {arXiv:1907.00952 [cond-mat.str-el]}
  \BibitemShut {NoStop}%
\bibitem [{\citenamefont {{Kwan}}\ \emph {et~al.}(2021)\citenamefont {{Kwan}},
  \citenamefont {{Wagner}}, \citenamefont {{Soejima}}, \citenamefont
  {{Zaletel}}, \citenamefont {{Simon}}, \citenamefont {{Parameswaran}},\ and\
  \citenamefont {{Bultinck}}}]{kwan2021kekule}%
  \BibitemOpen
  \bibfield  {author} {\bibinfo {author} {\bibfnamefont {Y.~H.}\ \bibnamefont
  {{Kwan}}}, \bibinfo {author} {\bibfnamefont {G.}~\bibnamefont {{Wagner}}},
  \bibinfo {author} {\bibfnamefont {T.}~\bibnamefont {{Soejima}}}, \bibinfo
  {author} {\bibfnamefont {M.~P.}\ \bibnamefont {{Zaletel}}}, \bibinfo {author}
  {\bibfnamefont {S.~H.}\ \bibnamefont {{Simon}}}, \bibinfo {author}
  {\bibfnamefont {S.~A.}\ \bibnamefont {{Parameswaran}}},\ and\ \bibinfo
  {author} {\bibfnamefont {N.}~\bibnamefont {{Bultinck}}},\ }\bibfield  {title}
  {\bibinfo {title} {{Kekul{\'e} Spiral Order at All Nonzero Integer Fillings
  in Twisted Bilayer Graphene}},\ }\href
  {https://doi.org/10.1103/PhysRevX.11.041063} {\bibfield  {journal} {\bibinfo
  {journal} {Physical Review X}\ }\textbf {\bibinfo {volume} {11}},\ \bibinfo
  {eid} {041063} (\bibinfo {year} {2021})},\ \Eprint
  {https://arxiv.org/abs/2105.05857} {arXiv:2105.05857 [cond-mat.str-el]}
  \BibitemShut {NoStop}%
\bibitem [{\citenamefont {{Wang}}\ \emph {et~al.}(2023)\citenamefont {{Wang}},
  \citenamefont {{Parker}}, \citenamefont {{Soejima}}, \citenamefont
  {{Hauschild}}, \citenamefont {{Anand}}, \citenamefont {{Bultinck}},\ and\
  \citenamefont {{Zaletel}}}]{wang2023_dmrgiks}%
  \BibitemOpen
  \bibfield  {author} {\bibinfo {author} {\bibfnamefont {T.}~\bibnamefont
  {{Wang}}}, \bibinfo {author} {\bibfnamefont {D.~E.}\ \bibnamefont
  {{Parker}}}, \bibinfo {author} {\bibfnamefont {T.}~\bibnamefont {{Soejima}}},
  \bibinfo {author} {\bibfnamefont {J.}~\bibnamefont {{Hauschild}}}, \bibinfo
  {author} {\bibfnamefont {S.}~\bibnamefont {{Anand}}}, \bibinfo {author}
  {\bibfnamefont {N.}~\bibnamefont {{Bultinck}}},\ and\ \bibinfo {author}
  {\bibfnamefont {M.~P.}\ \bibnamefont {{Zaletel}}},\ }\bibfield  {title}
  {\bibinfo {title} {{Ground-state order in magic-angle graphene at filling
  {\ensuremath{\nu}} ={\ensuremath{-}}3 : A full-scale density matrix
  renormalization group study}},\ }\href
  {https://doi.org/10.1103/PhysRevB.108.235128} {\bibfield  {journal} {\bibinfo
   {journal} {\prb}\ }\textbf {\bibinfo {volume} {108}},\ \bibinfo {eid}
  {235128} (\bibinfo {year} {2023})},\ \Eprint
  {https://arxiv.org/abs/2211.02693} {arXiv:2211.02693 [cond-mat.str-el]}
  \BibitemShut {NoStop}%
\bibitem [{\citenamefont {{Kim}}\ \emph {et~al.}(2023)\citenamefont {{Kim}},
  \citenamefont {{Choi}}, \citenamefont {{Lantagne-Hurtubise}}, \citenamefont
  {{Lewandowski}}, \citenamefont {{Thomson}}, \citenamefont {{Kong}},
  \citenamefont {{Zhou}}, \citenamefont {{Baum}}, \citenamefont {{Zhang}},
  \citenamefont {{Holleis}}, \citenamefont {{Watanabe}}, \citenamefont
  {{Taniguchi}}, \citenamefont {{Young}}, \citenamefont {{Alicea}},\ and\
  \citenamefont {{Nadj-Perge}}}]{kim2023_ivc}%
  \BibitemOpen
  \bibfield  {author} {\bibinfo {author} {\bibfnamefont {H.}~\bibnamefont
  {{Kim}}}, \bibinfo {author} {\bibfnamefont {Y.}~\bibnamefont {{Choi}}},
  \bibinfo {author} {\bibfnamefont {{\'E}.}~\bibnamefont
  {{Lantagne-Hurtubise}}}, \bibinfo {author} {\bibfnamefont {C.}~\bibnamefont
  {{Lewandowski}}}, \bibinfo {author} {\bibfnamefont {A.}~\bibnamefont
  {{Thomson}}}, \bibinfo {author} {\bibfnamefont {L.}~\bibnamefont {{Kong}}},
  \bibinfo {author} {\bibfnamefont {H.}~\bibnamefont {{Zhou}}}, \bibinfo
  {author} {\bibfnamefont {E.}~\bibnamefont {{Baum}}}, \bibinfo {author}
  {\bibfnamefont {Y.}~\bibnamefont {{Zhang}}}, \bibinfo {author} {\bibfnamefont
  {L.}~\bibnamefont {{Holleis}}}, \bibinfo {author} {\bibfnamefont
  {K.}~\bibnamefont {{Watanabe}}}, \bibinfo {author} {\bibfnamefont
  {T.}~\bibnamefont {{Taniguchi}}}, \bibinfo {author} {\bibfnamefont {A.~F.}\
  \bibnamefont {{Young}}}, \bibinfo {author} {\bibfnamefont {J.}~\bibnamefont
  {{Alicea}}},\ and\ \bibinfo {author} {\bibfnamefont {S.}~\bibnamefont
  {{Nadj-Perge}}},\ }\bibfield  {title} {\bibinfo {title} {{Imaging
  inter-valley coherent order in magic-angle twisted trilayer graphene}},\
  }\href {https://doi.org/10.1038/s41586-023-06663-8} {\bibfield  {journal}
  {\bibinfo  {journal} {\nat}\ }\textbf {\bibinfo {volume} {623}},\ \bibinfo
  {pages} {942} (\bibinfo {year} {2023})},\ \Eprint
  {https://arxiv.org/abs/2304.10586} {arXiv:2304.10586 [cond-mat.str-el]}
  \BibitemShut {NoStop}%
\bibitem [{\citenamefont {{Stepanov}}\ \emph {et~al.}(2020)\citenamefont
  {{Stepanov}}, \citenamefont {{Das}}, \citenamefont {{Lu}}, \citenamefont
  {{Fahimniya}}, \citenamefont {{Watanabe}}, \citenamefont {{Taniguchi}},
  \citenamefont {{Koppens}}, \citenamefont {{Lischner}}, \citenamefont
  {{Levitov}},\ and\ \citenamefont {{Efetov}}}]{stepanov2020_screening}%
  \BibitemOpen
  \bibfield  {author} {\bibinfo {author} {\bibfnamefont {P.}~\bibnamefont
  {{Stepanov}}}, \bibinfo {author} {\bibfnamefont {I.}~\bibnamefont {{Das}}},
  \bibinfo {author} {\bibfnamefont {X.}~\bibnamefont {{Lu}}}, \bibinfo {author}
  {\bibfnamefont {A.}~\bibnamefont {{Fahimniya}}}, \bibinfo {author}
  {\bibfnamefont {K.}~\bibnamefont {{Watanabe}}}, \bibinfo {author}
  {\bibfnamefont {T.}~\bibnamefont {{Taniguchi}}}, \bibinfo {author}
  {\bibfnamefont {F.~H.~L.}\ \bibnamefont {{Koppens}}}, \bibinfo {author}
  {\bibfnamefont {J.}~\bibnamefont {{Lischner}}}, \bibinfo {author}
  {\bibfnamefont {L.}~\bibnamefont {{Levitov}}},\ and\ \bibinfo {author}
  {\bibfnamefont {D.~K.}\ \bibnamefont {{Efetov}}},\ }\bibfield  {title}
  {\bibinfo {title} {{Untying the insulating and superconducting orders in
  magic-angle graphene}},\ }\href {https://doi.org/10.1038/s41586-020-2459-6}
  {\bibfield  {journal} {\bibinfo  {journal} {\nat}\ }\textbf {\bibinfo
  {volume} {583}},\ \bibinfo {pages} {375} (\bibinfo {year} {2020})},\ \Eprint
  {https://arxiv.org/abs/1911.09198} {arXiv:1911.09198 [cond-mat.supr-con]}
  \BibitemShut {NoStop}%
\bibitem [{\citenamefont {{Saito}}\ \emph {et~al.}(2020)\citenamefont
  {{Saito}}, \citenamefont {{Ge}}, \citenamefont {{Watanabe}}, \citenamefont
  {{Taniguchi}},\ and\ \citenamefont {{Young}}}]{saito2020_screening}%
  \BibitemOpen
  \bibfield  {author} {\bibinfo {author} {\bibfnamefont {Y.}~\bibnamefont
  {{Saito}}}, \bibinfo {author} {\bibfnamefont {J.}~\bibnamefont {{Ge}}},
  \bibinfo {author} {\bibfnamefont {K.}~\bibnamefont {{Watanabe}}}, \bibinfo
  {author} {\bibfnamefont {T.}~\bibnamefont {{Taniguchi}}},\ and\ \bibinfo
  {author} {\bibfnamefont {A.~F.}\ \bibnamefont {{Young}}},\ }\bibfield
  {title} {\bibinfo {title} {{Independent superconductors and correlated
  insulators in twisted bilayer graphene}},\ }\href
  {https://doi.org/10.1038/s41567-020-0928-3} {\bibfield  {journal} {\bibinfo
  {journal} {Nature Physics}\ }\textbf {\bibinfo {volume} {16}},\ \bibinfo
  {pages} {926} (\bibinfo {year} {2020})},\ \Eprint
  {https://arxiv.org/abs/1911.13302} {arXiv:1911.13302 [cond-mat.mes-hall]}
  \BibitemShut {NoStop}%
\bibitem [{\citenamefont {{Liu}}\ \emph {et~al.}(2021)\citenamefont {{Liu}},
  \citenamefont {{Wang}}, \citenamefont {{Watanabe}}, \citenamefont
  {{Taniguchi}}, \citenamefont {{Vafek}},\ and\ \citenamefont
  {{Li}}}]{liu2021_screening}%
  \BibitemOpen
  \bibfield  {author} {\bibinfo {author} {\bibfnamefont {X.}~\bibnamefont
  {{Liu}}}, \bibinfo {author} {\bibfnamefont {Z.}~\bibnamefont {{Wang}}},
  \bibinfo {author} {\bibfnamefont {K.}~\bibnamefont {{Watanabe}}}, \bibinfo
  {author} {\bibfnamefont {T.}~\bibnamefont {{Taniguchi}}}, \bibinfo {author}
  {\bibfnamefont {O.}~\bibnamefont {{Vafek}}},\ and\ \bibinfo {author}
  {\bibfnamefont {J.~I.~A.}\ \bibnamefont {{Li}}},\ }\bibfield  {title}
  {\bibinfo {title} {{Tuning electron correlation in magic-angle twisted
  bilayer graphene using Coulomb screening}},\ }\href
  {https://doi.org/10.1126/science.abb8754} {\bibfield  {journal} {\bibinfo
  {journal} {Science}\ }\textbf {\bibinfo {volume} {371}},\ \bibinfo {pages}
  {1261} (\bibinfo {year} {2021})},\ \Eprint {https://arxiv.org/abs/2003.11072}
  {arXiv:2003.11072 [cond-mat.mes-hall]} \BibitemShut {NoStop}%
\end{thebibliography}%

\end{document}